\documentclass[aps,prl,twocolumn,reprint, showpacs]{revtex4-1}
\pdfoutput=1

\usepackage[sort&compress]{natbib}
\usepackage{amsfonts}
\usepackage[cmex10]{amsmath}
\usepackage{bm}
\usepackage{color}
\usepackage{soul}
\sethlcolor{white}

\usepackage[]{caption}
\usepackage[font=footnotesize]{subfig}

\usepackage[pdftex]{graphicx}
\usepackage{lipsum}

\begin{document}

\title{Design of Tensor Impedance Transmitarrays For Polarization Control}

\author{Michael~Selvanayagam and George.~V.~Eleftheriades }
\affiliation{The Edward S. Rogers Department
of Electrical and Computer Engineering, University of Toronto, Toronto}

\date{\today}



\begin{abstract}
Tensor impedance transmitarrays consist of layers of  tensor impedance surfaces, separated by dielectric spacers.  Because the surface impedance of each layer is a tensor, an arbitrarily polarized  incident field is scattered into its constituent TE and TM components. This gives rise to transmitarrays capable of altering the polarization state of an incident field.  Various  tensor impedance transmitarray  have been proposed in the literature to alter the polarization, however no comprehensive methodology has been proposed to design these structures.

In this work we propose a procedure for designing tensor impedance transmitarrays using multi-conductor transmission-line (MTL) theory.   We treat the transmitarray by modelling free-space and the dielectric spacers as an MTL supporting TE and TM modes with each tensor impedance surface as a shunt load along the MTL. By using simple MTL concepts we can design a transmitarray to be reflectionless while controlling the transmission through the layers and thus the transmitted polarization state.

We demonstrate this procedure for two classes of tensor impedance transmitarrays while also validating the design using full-wave simulation. The first class are symmetric transmitarrays which can alter a given incident polarization state into a desired polarization state.  The second class are asymmetric transmitarrays which can also implement chiral polarization effects such as a linear polarization rotation and a circular polarization selectivity.

\end{abstract}

\maketitle


\section{Introduction}
\label{sec:intro}
Transmitarrays are a class of antenna arrays capable of providing high-gain apertures to generate directive radiation for communication links and imaging applications \cite{McGrath_1986,Pozar_1996}.  These arrays work to alter the incident wavefront by providing the desired phase-shift across the aperture itself.  Transmitarrays can implement this phase-shifting behaviour using different architectures.  One common architecture is to use back-to-back  configurations of antennas to receive  and retransmit  the incident field \cite{Pozar_1996,Popovic_1998}. Here, phase shifting circuitry is included between the antennas to control the phase shift imparted 	by the array \cite{MunozAcevedo_2009}.  These designs can also be made reconfigurable \cite{Lau_2011,Lau_Hum_2012,Hum_2014}.

Another common configuration are stacks of passive scatterers \cite{Ryan_etal_2010,AlJoymayly2011,Li_2013,Gagnon_2013}. Each unit cell in the stack is individually tuned to provide a desired phase shift across the aperture with unity transmission and no reflection. These stacks of scatterers can be interpreted as frequency selective surfaces or impedance sheets \cite{AlJoymayly2011,Gagnon_2013}.

While all of these examples utilized the microwave portion of the electromagnetic spectrum, the concept of a transmitarray has been recently extended to the terahertz and photonic portion of the spectrum. These include terahertz and optical versions of scatterer based transmitarrays \cite{Memarzadeh_2011,Monticone_etal_2013} as well as single layer structures which have limited performance \cite{Yu_etal_2011,Kildishev_etal_2013}.

Another class of related devices that can control wavefronts are metasurfaces which use elementary dipole elements as  scatterers on a surface\cite{Holloway_etal_2005,Holloway_etal_2012,Vehmas_2013}.  One example of this is the Huygens surface which consists of a superposition of electric and magnetic dipoles.  This design can maximize transmission through a single surface while providing wavefront control \cite{Pfeiffer_Grbic_2013,Selvanayagam_Eleftheriades_2013}.

While wavefront control has been successfully demonstrated using various transmitarray architectures, controlling other aspects of the electromagnetic field using a transmitarray is an active area of research. One of the main areas of investigation is the polarization of the electromagnetic field and how it can be controlled using transmitarrays.  

At microwave frequencies, various examples include transmitarrays that can provide phase control to incident TE and TM waves using a stack of scatterers \cite{Strassner_2004,Joyal_2012}. Transmitarrays constructed from back-to-back antennas can also provide polarization control using antennas that have been rotated to alter the polarzation of the field radiated by each antenna \cite{Nakatani_1977,Kaouach2011, Phillion_2011}. 

At both microwave and optical frequencies various polarization controlling devices have been proposed using metasurfaces.  This includes devices which can act like quarter-wave and half-wave plates \cite{Zhu_2013,Pfeiffer_2013_2,Farmahini_2013,Yu_2012,He_2013,Pfeiffer_2013,Grady2013,Yang2014} and Huygens surfaces that can alter the polarization state for a specifically designed incident field \cite{Selvanayagam2014}. 

Another way of altering the polarization involves using surfaces of bianisotropic particles which includes chiral particles as a subset. Bianisotropic and chiral particles/surfaces that have been designed to alter the polarization of incident plane wave have been proposed in \cite{Papakostas2003,Rogacheva2006,Decker2007,Soukoulis2011,Niemi2013}. With regards to chiral behaviour, impedance surfaces have also been proposed as a way manifesting chirality \cite{Plum2009,Zhao_etal_2011,Zhao_2012,Joyal2012}.

For many of these polarization controlling metasurfaces, the surfaces can be described by tensor surface impedances.  This is due to the presence of structures such as rotated dipoles and wires \cite{Zhao_2012,Grady2013} , `V'-antennas \cite{Yu_2012} and silicon resonators \cite{Yang2014}. Also, many of these designs consist of stacks of  metasurfaces separated by dielectric spacers for minimizing diffraction orders and/or unwanted reflections from the surface \cite{Zhao_2012,Yang2014}. However no consistent methodology has been proposed which can exactly specify the appropriate tensor impedance of each layer to achieve a desired polarization function with a reflectionless transmitarray.


This raises the question of how one can design a tensor impedance transmitarray with the appropriate impedances on each layer to achieve the desired reflection and transmission coefficients.   In this work we tackle this problem by proposing a multi-conductor transmission-line (MTL) model of a general tensor impedance transmitarray to understand how waves are reflected and transmitted through such a structure. We will then look at how this model can be used to design two types of transmitarrays. The first kind are symmetric tensor impedance transmitarrays capable of implementing wave-plate type polarization control as well as inhomogeneous polarization control across the aperture of the transmitarray. The second kind are asymmetric tensor impedance transmitarrays which have the added capability of implementing chiral polarization behaviour such as polarization rotation and circular polarization selectivity. Throughout this process we will demonstrate our design procedure using various examples verified with full-wave simulations.



\section{Theory}
\label{sec:Theory}
\begin{figure}[!t]
\centering
\subfloat[]
{
	\includegraphics[clip=true, trim= 0cm 2cm 0cm 1cm, scale=0.25]{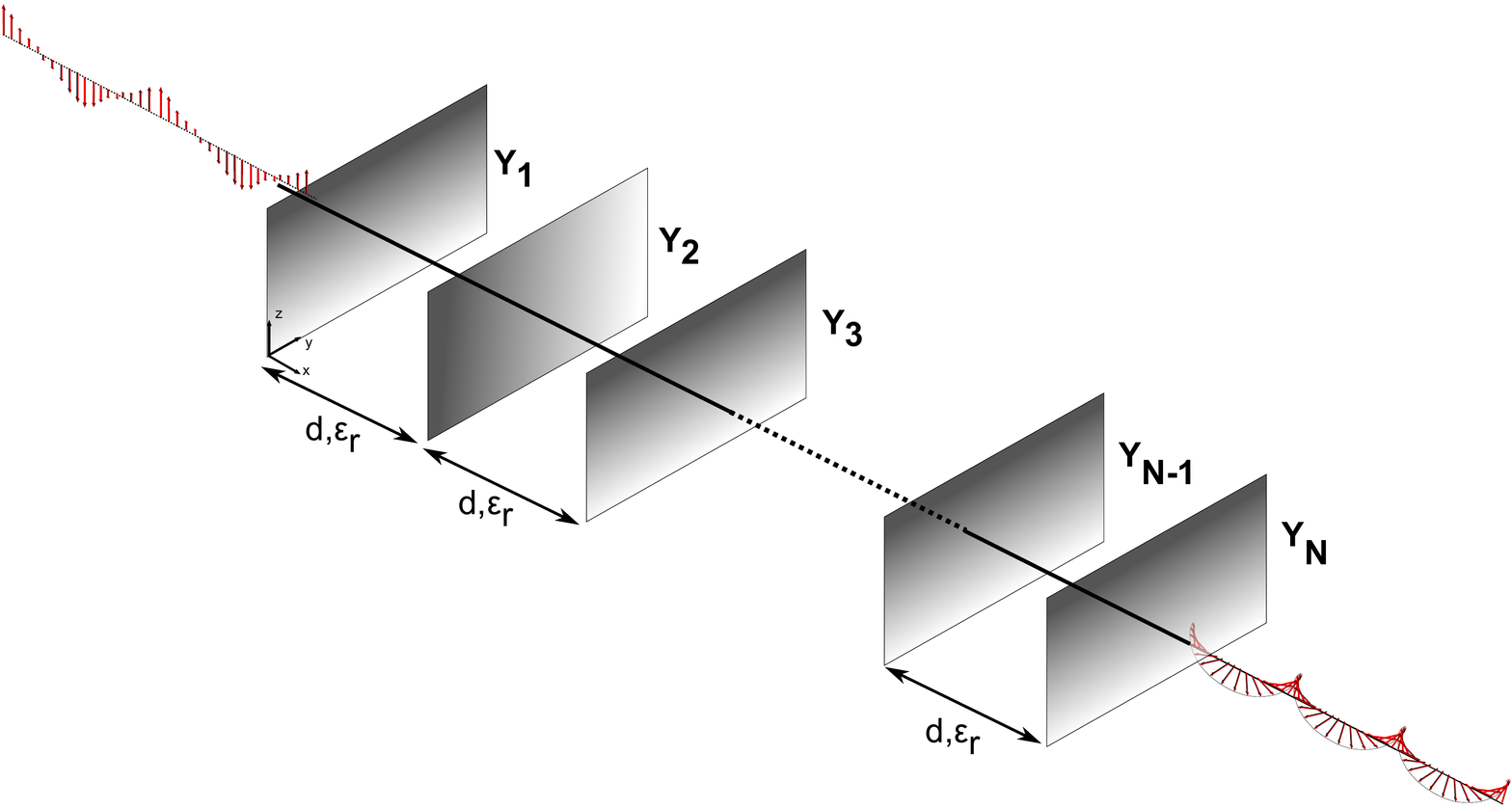}
	\label{fig:TensorTransmitarraySchemWave}
}
\\
\subfloat[]
{
	\includegraphics[clip=true, trim=  0cm 5cm 1cm 1cm, scale=0.25]{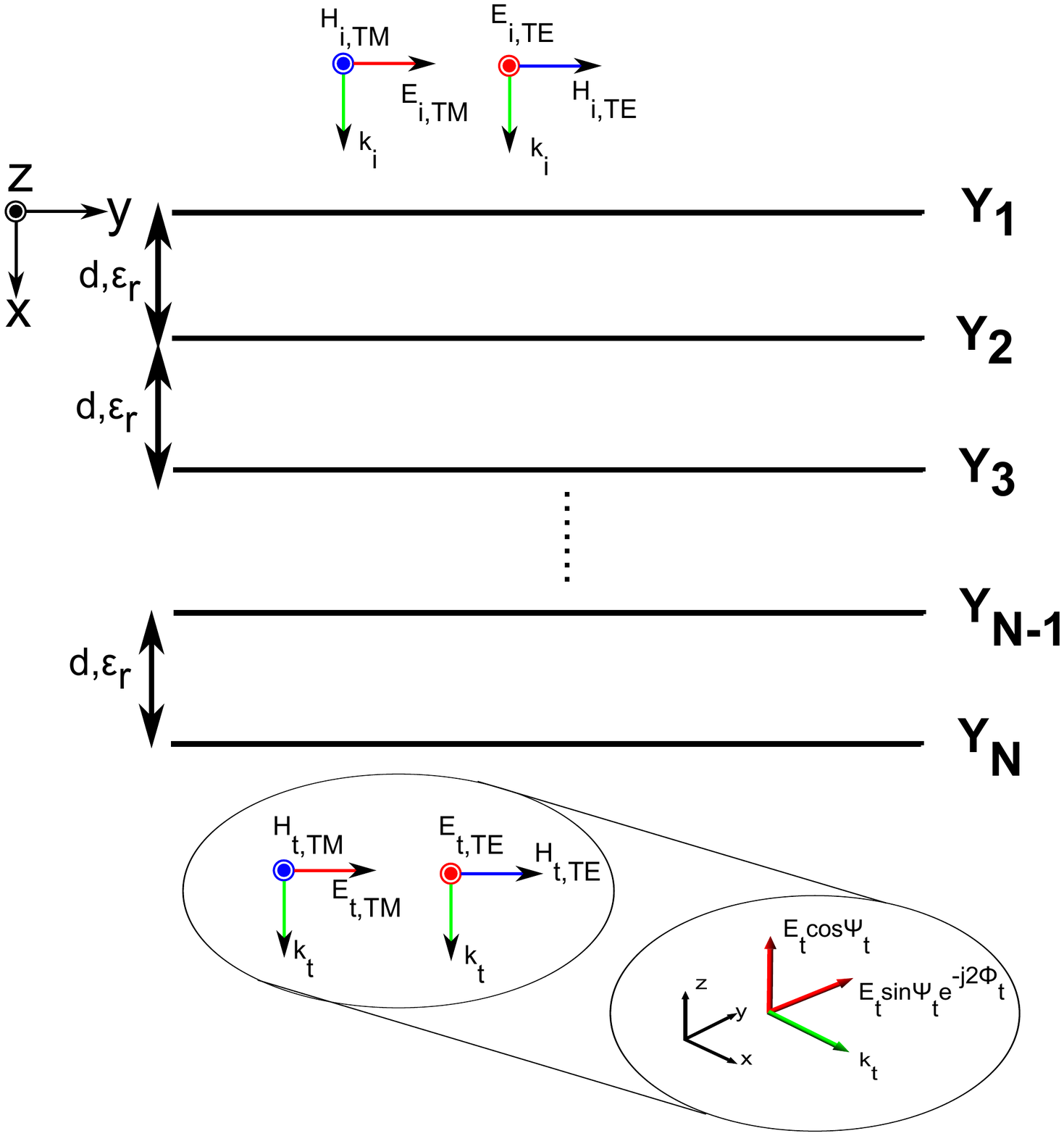}
	\label{fig:TensorTransmitarrayFields}
}

\caption{\protect\subref{fig:TensorTransmitarraySchemWave} A schematic of an $N$ layer tensor impedance transmitarray.  Each sheet of the transmitarray consists of a tensor admittance.   \protect\subref{fig:TensorTransmitarrayFields} The fields on either side of the transmitarray as decomposed into TE and TM modes.}
\label{fig:TensorTransmitarray}
\end{figure}

A tensor impedance transmitarray consists of multiple impedance surfaces stacked on top of each other and separated by a dielectric spacer. This is pictured in Fig.~\ref{fig:TensorTransmitarraySchemWave}.  Our coordinate system is also defined in Fig.~\ref{fig:TensorTransmitarraySchemWave}.  Here, each impedance surface sits on a plane parallel to the $yz$-plane starting at $x=0$ separated by a dielectric spacer with equal spacing $d$ and dielectric constant $\varepsilon_r$. We will assume that the spacers are identical between each surface, though this assumption can easily be relaxed as needed. 

To simplify the problem we look at plane-wave propagation in the $xy$-plane through this stack of impedance sheets, with propagation confined to the $x$-axis.  In general, the incident and transmitted waves can enter at any angle as such a surface is also capable of beamshaping, however this is outside the scope of this work \cite{Monticone_etal_2013,Pfeiffer_2013_2}. However as we will see it is easy to incorporate this functionality once we have a suitable reflection and transmission model for the transmitarray

The normal plane-waves travelling through the transmitarray can be described by a linear combination of  TE and TM waves.  Here we define a TE wave to have an electric field transverse to  the $xy$-plane of propagation and a TM wave defined to have a magnetic field transverse to the $xy$-plane of propagation (or an in-plane electric field) as shown in Fig.~\ref{fig:TensorTransmitarrayFields}. 

To succinctly describe the polarization state of the incident and transmitted field on either side of the transmitarray we introduce two variables, $\Psi$ and $\Phi$.  As shown in Fig.~\ref{fig:TensorTransmitarrayFields}, $\Psi$ describes the relative amplitude between the TE and TM fields while $\Phi$ describes the relative phase.  We can see that these variables correspond to the definition of various polarization states on the unit Poincare sphere \cite{Born_Wolf}.

The unique aspect of the transmitarray under discussion is that each impedance surface is described by a tensor impedance or admittance. This tensor impedance relates continuous electric field at the boundary to the discontinuity in the magnetic field as given by 
\begin{equation}
\label{eq:YtensorBoundary}
\left[\begin{array}{c} E_y \\ E_z \end{array} \right]= 
\left[\begin{array}{cc} Y_{yy} & Y_{yz} \\ Y_{yz} & Y_{zz} \end{array} \right]^{-1}\left[\begin{array}{c} -(H_{z,2}-H_{z,1}) \\H_{y,2}-H_{y,1} \end{array} \right].
\end{equation}
This is different from traditional transmitarray architectures where each surface is described by a scalar (or isotropic) impedance/admittance, which we will refer to as scalar impedance transmitarrays. From here on we write and refer to the quantity in \eqref{eq:YtensorBoundary} as an admittance tensor (and admittance surface) for reasons that will become clear below. 

\begin{figure}[!t]
\centering
\includegraphics[clip=true, trim= 0cm 0cm 0cm 0cm, scale=0.25]{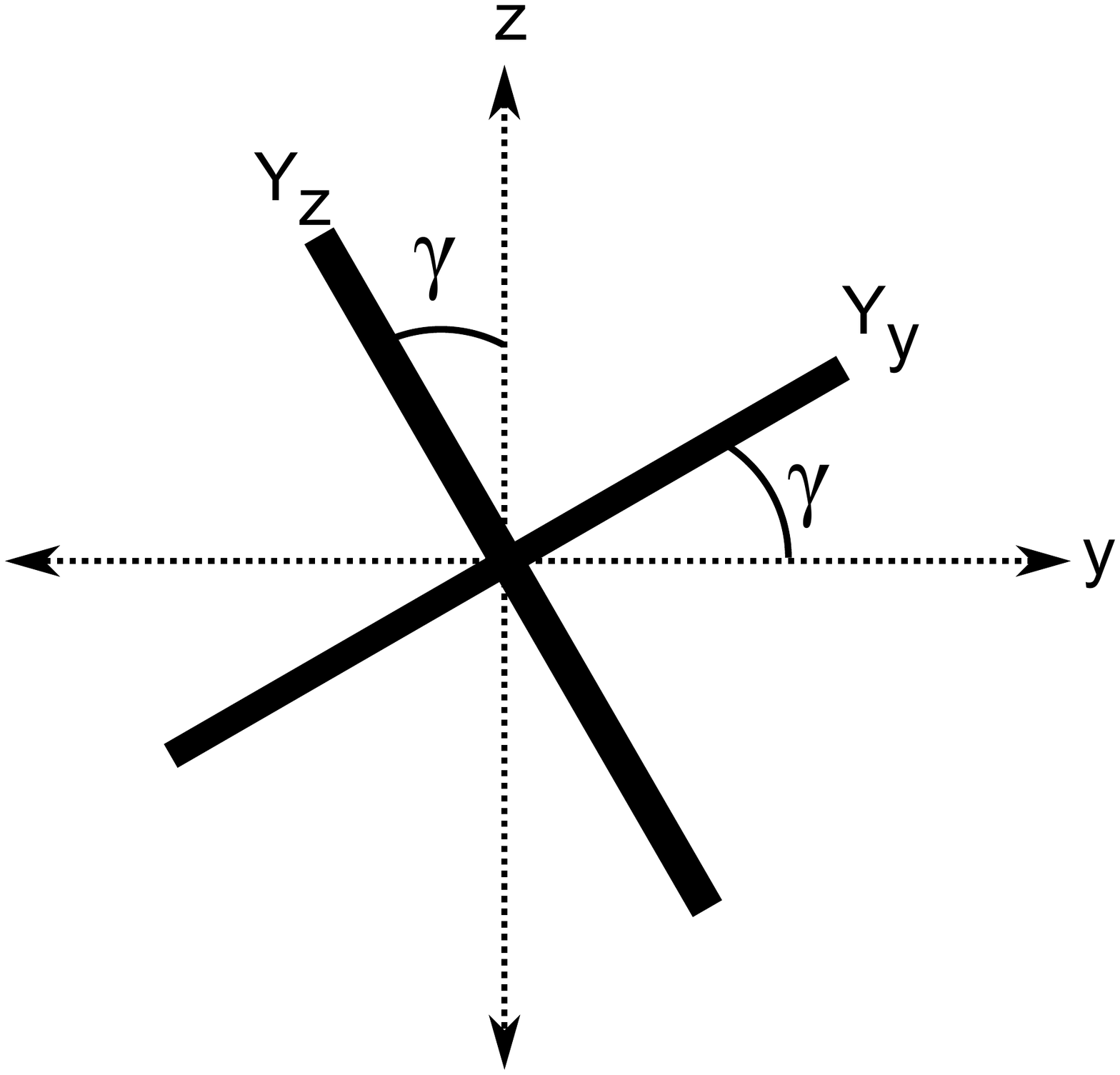}
\caption{A physical interpretation of a diagonalized tensor admittance surface. The admittance of each arm of the crossed dipole corresponds to the eigenvalues and the rotation angle $\gamma$ to the rotation matrix that diagonalizes the admittance tensor. }
\label{fig:CrossedDipoleEigenvalue}
\end{figure}

This admittance tensor is imaginary and anti-Hermitian to satisfy energy conservation and reciprocity  \cite{Fong_etal_2010}.  It can be seen then that this tensor has three degrees of freedom which can be most easily understood by diagonalizing the tensor \cite{Selvanayagam_Eleftheriades_2011_3,Patel_2013_2,Selvanayagam2014}. Doing this gives the following expression,
\begin{align}
\label{eq:Ydiag}
\mathbf{Y}=\mathbf{R}(\gamma)  \left[\begin{array}{cc} Y_{y} & 0\\0 & Y_{z} \end{array}  \right] \mathbf{R}^{-1}(\gamma),
\end{align}
where the diagonalization matrix is a rotation matrix, characterized by an angle $\gamma$ and given by,
\begin{equation}
\mathbf{R}(\gamma)= \left[\begin{array}{cc} \cos\gamma & -\sin\gamma \\ \sin\gamma & \cos\gamma \end{array}  \right].
\end{equation}
The simplest interpretation of this is to map \eqref{eq:Ydiag} to the geometry of a crossed dipole as shown in Fig.~\ref{fig:CrossedDipoleEigenvalue}.  Here the eigenvalues map to the surface admittance of each arm of the crossed dipole, while $\gamma$ corresponds to the rotation angle of the entire crossed dipole. 

As we will see below, by varying these three degrees of freedom we can alter the surface admittance to design the transmitarray.   We also note that there are many ways to implement a surface admittance besides a crossed dipole as we will discuss in Section~\ref{sec:Conclusion} and that this physical interpretation is just a convenient way to understand the admittance tensor.

Given that each admittance surface in the transmitarray is described by a tensor admittance it is clear that a TE-polarized plane-wave can couple to a TM-polarized plane wave and vice versa. Thus it is conceivable that by properly tuning the admittance of each layer we can:
\begin{enumerate}
\item Eliminate (or control) reflections.
\item Control the relative phase between the transmitted TE and TM polarized fields ($\Phi_t$)
\item Control the relative amplitude between the transmitted TE and TM polarized fields ($\Psi_t$).
\item Control the absolute phase for beamshaping (which we ignore here for simplicity).
\end{enumerate}

To achieve this critera we need a model which allows us to specify the admittance of each layer of the transmitarray to achieve the stated goals.

\subsection{Multi-Conductor Transmission-Line Model}
The model that we will use to design a tensor impedance transmittary is based on multi-conductor transmission-line (MTL) theory. To understand why we choose this approach, we look at scalar impedance transmitarrays. For these kinds of transmitarrays, a simple two-wire transmission-line model is used, where each layer of the transmitarray  is modelled as a shunt impedance loading the transmission-line. By using some form of transmission-line analysis  such as filter theory \cite{AlJoymayly2011}, S-parameters analysis \cite{Monticone_etal_2013,Pfeiffer_2013_2} or Smith Chart design \cite{Lau_Hum_2012,Joyal_2012}, a suitable value for the scalar impedance of each layer can be found which minimizes reflections while varying the phase of the transmitted wave.

\begin{figure*}[!t]
\centering
\subfloat[]
{
	\includegraphics[clip=true, trim= 0cm 0cm 0cm 0cm, scale=0.2]{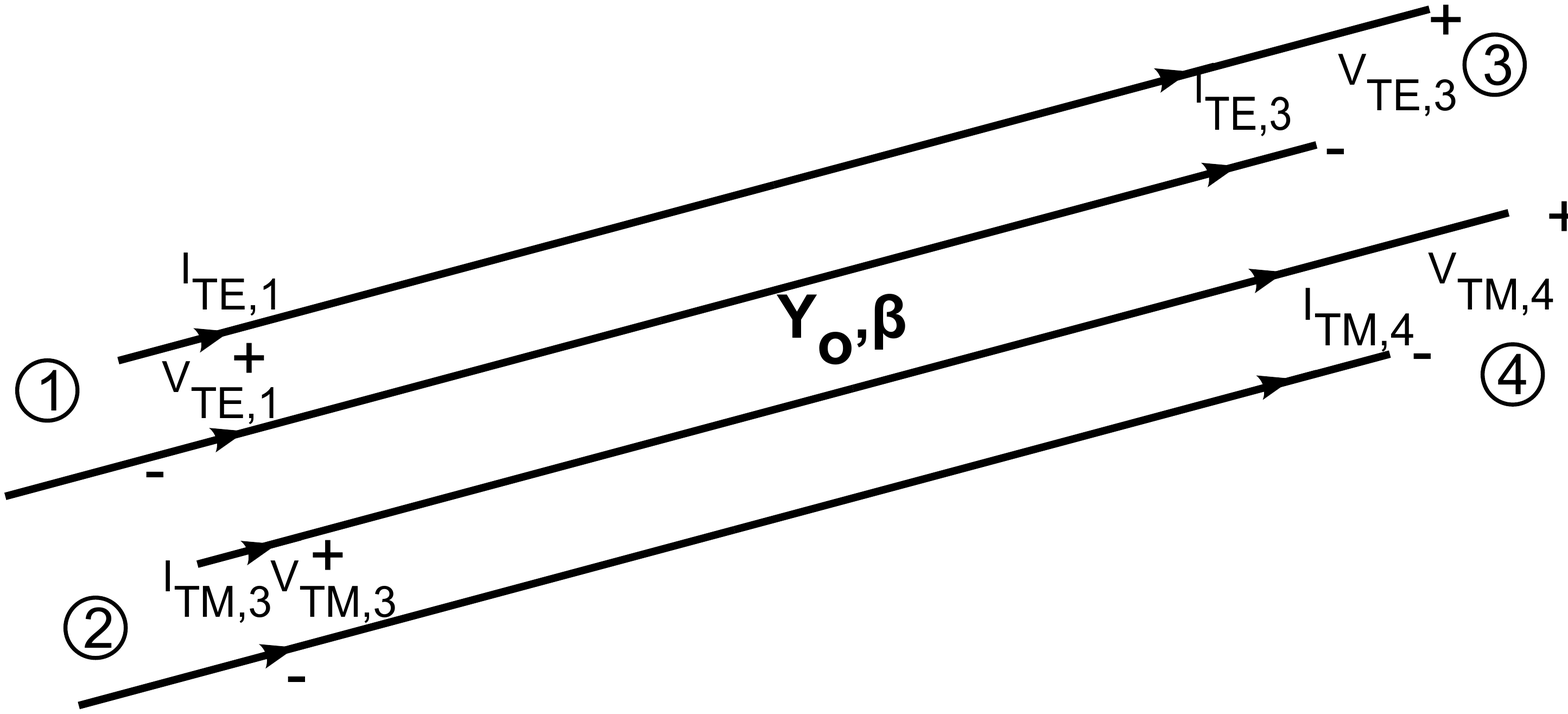}
	\label{fig:FreeSpaceMTL}
}
\subfloat[]
{
	\includegraphics[clip=true, trim=  0cm 0cm 0cm 0cm, scale=0.2]{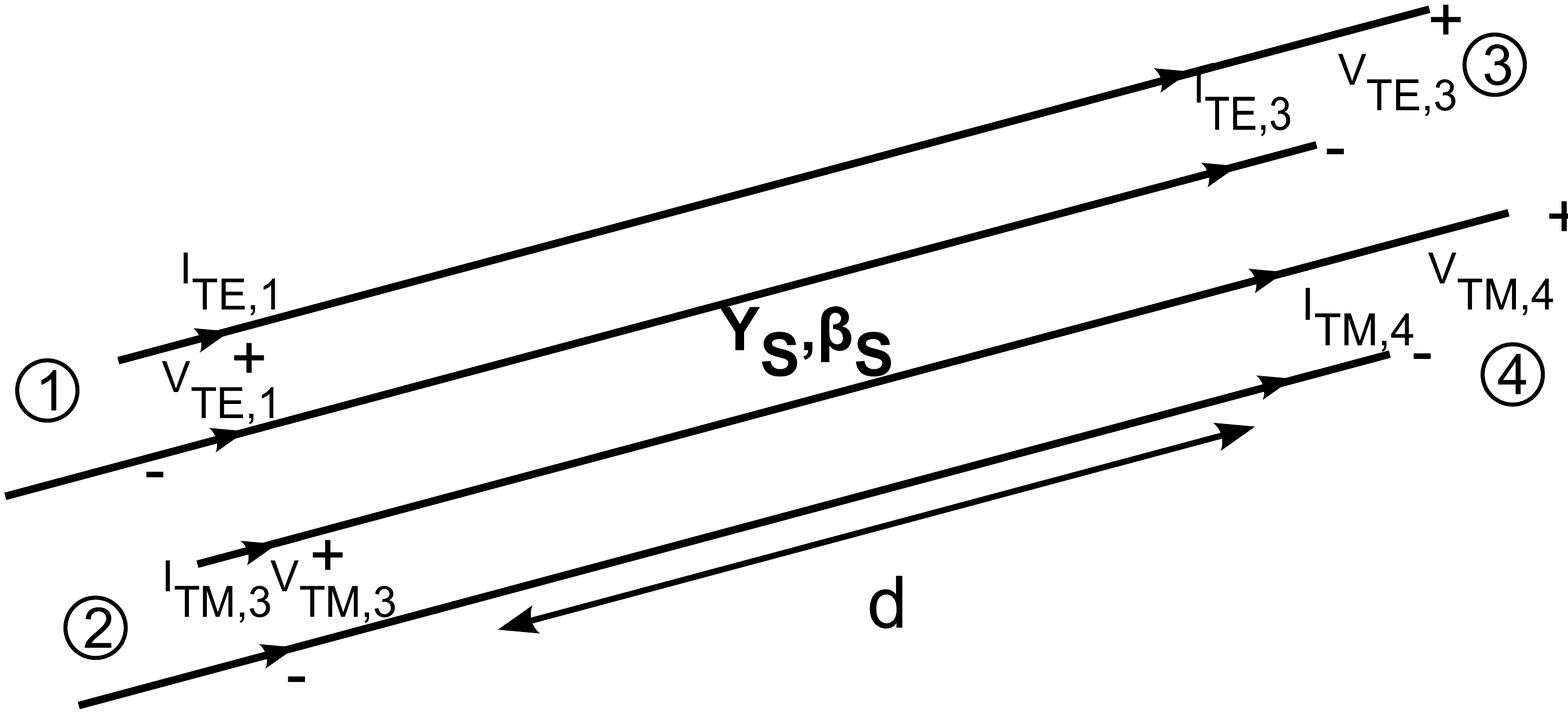}
	\label{fig:SpacerMTL}
}
\subfloat[]
{
	\includegraphics[clip=true, trim= 0cm 0cm 0cm 0cm, scale=0.2]{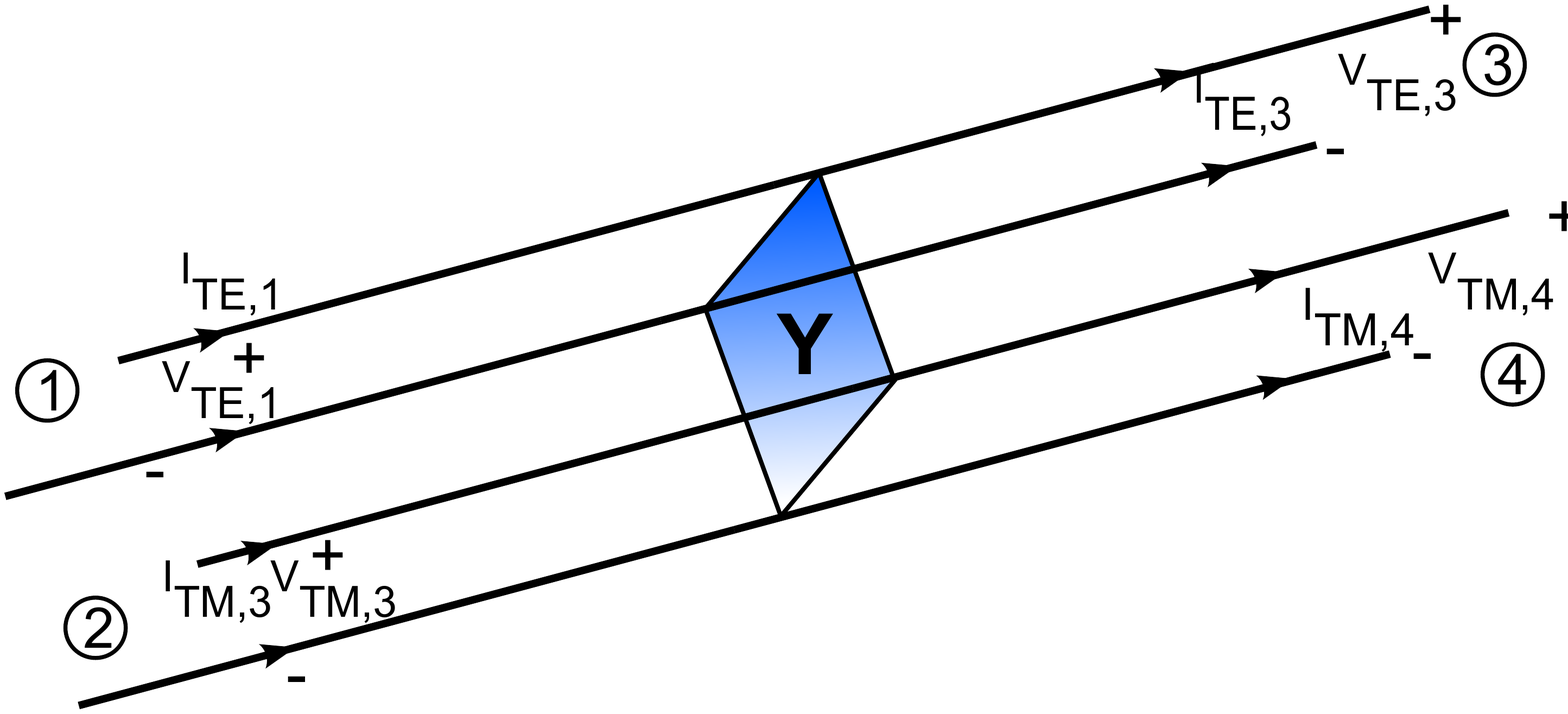}
	\label{fig:ShuntYMTL}
}
\caption{The three elements comprising the MTL model of a tensor admittance transmitarray\protect\subref{fig:FreeSpaceMTL} The MTL model of free space. The MTL supports a TE mode and TM mode as defined previously and as annotated above. The two modes are orthogonal in the MTL. Note that we repeat the ground wire for clarity. \protect\subref{fig:SpacerMTL}The MTL model of the dielectric spacer. \protect\subref{fig:ShuntYMTL} A shunt tensor admittance loading the MTL. This models the surface admittance of each layer of the transmitarray.}
\label{fig:TensorTransmitarray}
\end{figure*}

With this in mind we can extend some of these concepts from scalar transmitarrays to tensor transmitarrays by defining an appropriate MTL model since we have two modes supported by our structure.  We begin by looking at free-space surrounding the transmitarray. Because we are looking at one TE and one TM mode incident upon the transmitarray we can construct a simple $2+1$ wire MTL with one segment supporting a TE-polarized wave and the other supporting a TM-polarized wave. This is pictured in Fig.~\ref{fig:FreeSpaceMTL}. Note the numbering of the ports with the associated TE and TM waves on the MTL.  Because the TE and TM modes are orthogonal in free-space, the MTL model is as simple as it could possibly be, with the characteristic impedance matrix and propagation constant matrix being given by diagonal matrices,
\begin{align}
\pmb{\beta}&= \frac{2 \pi}{\lambda} \mathbf{E}, \\
\label{eq:CharImpMtx}
\mathbf{Y_o}&=\frac{1}{\eta}\mathbf{E},
\end{align} 
respectively where $\mathbf{E}$ is the identity matrix and $\eta$ is the impedance of free space.

This can also be done for the dielectric spacers of length $d$ as given in Fig.~\ref{fig:SpacerMTL} between each admittance surface where we define the corresponding MTL quantities to be
\begin{align}
\pmb{\beta_S}&= \frac{2 \pi \sqrt{\varepsilon_r}}{\lambda} \mathbf{E}, \\
\mathbf{Y_{s}}&=\frac{\sqrt{\varepsilon_r}}{\eta}\mathbf{E}.
\end{align}
Assuming the spacers are simple dielectrics we again have a basic MTL model. We note that these spacers should not be extremely subwavelength. Otherwise the coupling between each admittance surface dominates and the MTL model no longer holds since there are now multiple propagating and evanescent modes between each layer.  However, as demonstrated experimentally, these spacer layers can still be made compact \cite{Pfeiffer_2013_2,Zhao_2012} 

The last part of our model are the admittance surfaces between the spacers and these are described by \eqref{eq:YtensorBoundary} and shown in Fig.~\ref{fig:ShuntYMTL}. We can see, like their scalar transmitarray counterpart, that these elements are in shunt with the MTL transmission-line, hence our description of them as admittances. By cascading these elements together we form an MTL model of a tensor impedance transmitarray made up of $N$ layers.  

With this model, we can define a few basic quantities to characterize the fields within the transmitarray. At any point on the MTL line we can define the reflection coefficient matrix, $\mathbf{\Gamma}=\left[\begin{array}{cc} \Gamma_{yy} & \Gamma_{yz} \\ \Gamma_{zy} & \Gamma_{zz} \end{array} \right]$  and input admittance matrix $\mathbf{Y_{in}}=\left[\begin{array}{cc} Y_{in,yy} & Y_{in,yz} \\ Y_{in,zy} & Y_{in,zz} \end{array} \right]$. These are related to each other by,
\begin{align}
\label{eq:YfromGamma}
\mathbf{Y_{in}}=\mathbf{Y_o}(\mathbf{E}-\mathbf{\Gamma})(\mathbf{E}+\mathbf{\Gamma})^{-1}
\end{align}
\begin{align}
\label{eq:GammaFromY}
\mathbf{\Gamma}=(\mathbf{Y_o}-\mathbf{Y_L})(\mathbf{Y_o}+\mathbf{Y_L})^{-1}.
\end{align}

With these two quantities defined, the design methodology for a transmitarray is achieved by forcing the transmitarray to be reflectionless (or to have a specific reflection coefficient matrix). This means that we must force the reflection coefficient at the input of the transmitarray to $\mathbf{0}$ (matrix of zeros) and the input admittance at the input of the transmitarray has to be the same as \eqref{eq:CharImpMtx}. Once we have a reflectionless design, the transmission coefficients will fall into place. 

\begin{figure}[!t]
\centering
\includegraphics[clip=true, trim= 0cm 0cm 0cm 0cm, scale=0.25]{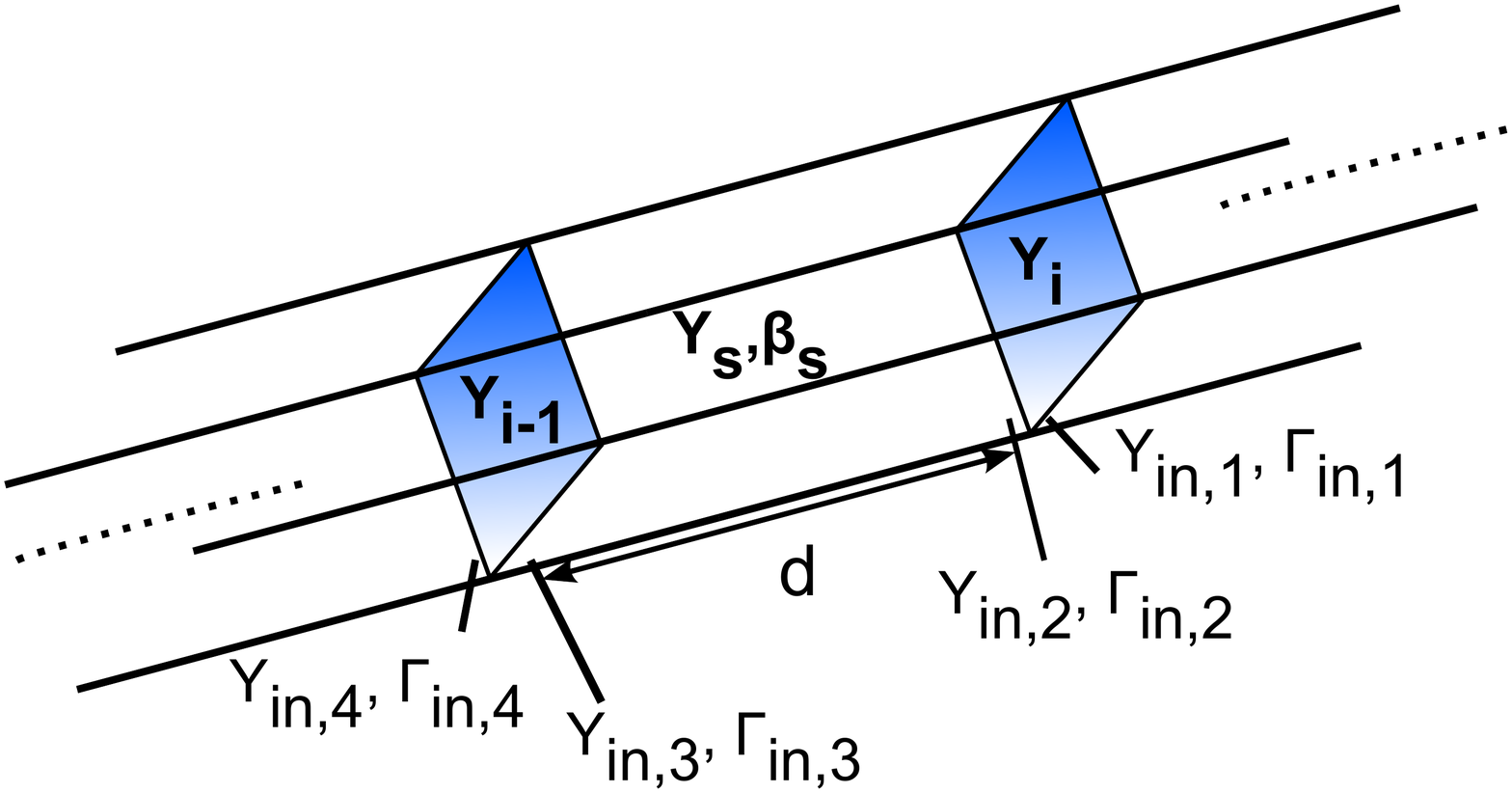}
\caption{A slice of an MTL model of a tensor impedance transmitarray between the $i$ and $i-1$ layers. There are two basic operations when examining the input admittance and reflection coefficient matrices from the $ith$ layer to the $i-1$ layer.}
\label{fig:MTLOps}
\end{figure}

To guarantee that the reflection coefficient and input admittance have the desired value at the input of the transmitarray we have at our disposal two basic MTL operations. These are
\begin{enumerate}
 \item The addition of a shunt admittance tensor.
 \item Translating an input admittance along an MTL.
 \end{enumerate}  
This is defined more concretely in Fig.~\ref{fig:MTLOps} where we take an arbitrary slice of an $N$ layer transmitarray. At point $1$ in this slice of the transmitarray as shown in Fig.~\ref{fig:MTLOps}, the point closest to the output, we have the input admittance matrix $\mathbf{Y_{in,1}}$ and the reflection coefficient matrix $\mathbf{\Gamma_{in,1}}$. The two MTL operations given above are the effect of the dielectric and surface admittance sheets on the input admittance and the reflection coefficient.

To see the effect of these two operations on the input admittance and the reflection coefficient matrices we can immediately look at the next point closest to the input of the transmitarray, point $2$, in Fig.~\ref{fig:MTLOps}. Here we see that the input admittance and reflection coefficient matricies change due to the shunt admittance tensor of the $i$'th surface. At this point the new input admittance is given by,
\begin{align}
\label{eq:Yi}
\mathbf{Y_{in,2}}=\mathbf{Y_{in,1}}-\mathbf{Y_i},
\end{align}
and the new reflection coefficient by substituting \eqref{eq:Yi} into \eqref{eq:GammaFromY}. This operation is capable of altering the imaginary part of the input admittance only. Moving to the next closest point to the input, point $3$, the input admittance and reflection coefficient matrices have been altered due to the dielectric spacer. In the MTL model this alters the reflection coefficient by \footnote{Strictly speaking the reflection coefficient is given by $\pmb{\Gamma}(d)=e^{-j\pmb{\beta} d } \pmb{\Gamma}(0) (e^{j\pmb{\beta}d})^{-1}$ where the matrix exponential operation is defined to be $e^{j\pmb{\beta} d}=\mathbf{P^{-1}}\left[\begin{array}{cc} e^{j \beta_1 d} & 0 \\ 0 & e^{j \beta_2 d} \end{array} \right]\mathbf{P}$, and $\mathbf{P}$ is the matrix that diagonalizes $e^{j\pmb{\beta}d}$. However since the TE and TM modes of free space and the dielectric spacer are orthogonal and isotropic this reduces to simply a scalar multiplication of the reflection coefficient matrix by a complex exponential. A similar statement applies to the $\tan{\beta d}$ term in (12).}
\begin{align}
\label{eq:GammaTL}
\mathbf{\Gamma_{in,3}}=e^{-j2 \beta d} \mathbf{\Gamma_{in,2}},
\end{align}
where $\beta=2\pi/\lambda$. Correspondingly the input admittance is given by
\begin{align}
\label{eq:YinTL}
\mathbf{Y_{in,3}}=\frac{1}{\eta} \left( \mathbf{Y_{in,2}}+ j \mathbf{Yo} \tan{\beta d} \right) \left( \mathbf{Y_{o}}+ j \mathbf{Y_{in,2}} \tan{\beta d} \right) ^{-1}.
\end{align}
This admittance translation operation alters both the real and imaginary parts of the input admittance and is our only degree of freedom to alter the real part.

As we continue to move through each layer of the transmitarray towards the input we keep on repeating these two operations for each of the $N$ surface admittance layers and the $N-1$ dielectric spacers of the transmitarray.  

With these two operations we have the tools we need to formulate a design procedure to construct reflectionless polariztaion controlling transmitarrays. In the rest of this paper we look at two classes of tensor impedance transmitarrays, symmetric and asymmetric transmitarrays. 

Symmetric transmitarrays are transmitarrays whose surface impedance is identical on the $1^{st}$ and $N^{th}$ layer, $2^{nd}$ and $(N-1)^{th}$, and so on. Similar to scalar impedance transmitarrays, to formulate a reflectionless transmitarray we only need $N=3$ layers  \cite{Monticone_etal_2013}. In section~\ref{sec:SymTxArray} we will discuss how to design these transmitarrays using the MTL operations given above.

Asymmetric transmitarrays have each layer different than the other and hence no symmetry around the center layer of the transmitarray. These will be used to design transmitarrays which can implement chiral behaviour such as polarization rotation. Because of the lack of symmetry in the structure we will need $N=4$ layers to design reflectionless polarization controlling transmitarrays. We will discuss the design of these structures in Section.~\ref{sec:AsymTxArray}.

Whether dealing with symmetric or asymmetric transmitarrays we need to analyze the transmission through a set of reflectionless surface admittance layers to be able to satisfy the remaining design criteria given above. This is done by finding the S-parameters of the transmitarray using 4-port transfer matrices. This is described in Appendix A for completeness sake. Once we are able to find the transmission through the transmitarray we can look through all the reflectionless solutions to find one with the appropriate transmission phase and amplitude for the TE and TM modes for a specific solution.

We end this section on a final note comparing our MTL model to the transmission-line model that was proposed in \cite{Zhao_etal_2011,Zhao_2012} to analyze a stack of impedance surfaces.  The key difference here is that in \cite{Zhao_etal_2011,Zhao_2012} the transmission-line model was used to analyze how changing $\gamma$ in the surface admittance tensor affected the transmission through the structure. As we will see in the coming sections, by explicitly defining our model as an MTL and using related concepts such as the input admittance and reflection coefficient matrices, we will be able to use all variables of the surface admittance (eigenvalues and rotation angles) to construct designs that are reflectionless.

\section{Symmetric Tensor Impedance Transmitarrays}
\label{sec:SymTxArray}

\begin{figure}[!t]
\centering
\subfloat[]
{
	\includegraphics[clip=true, trim= 0cm 4cm 0cm 3.5cm, scale=0.22]{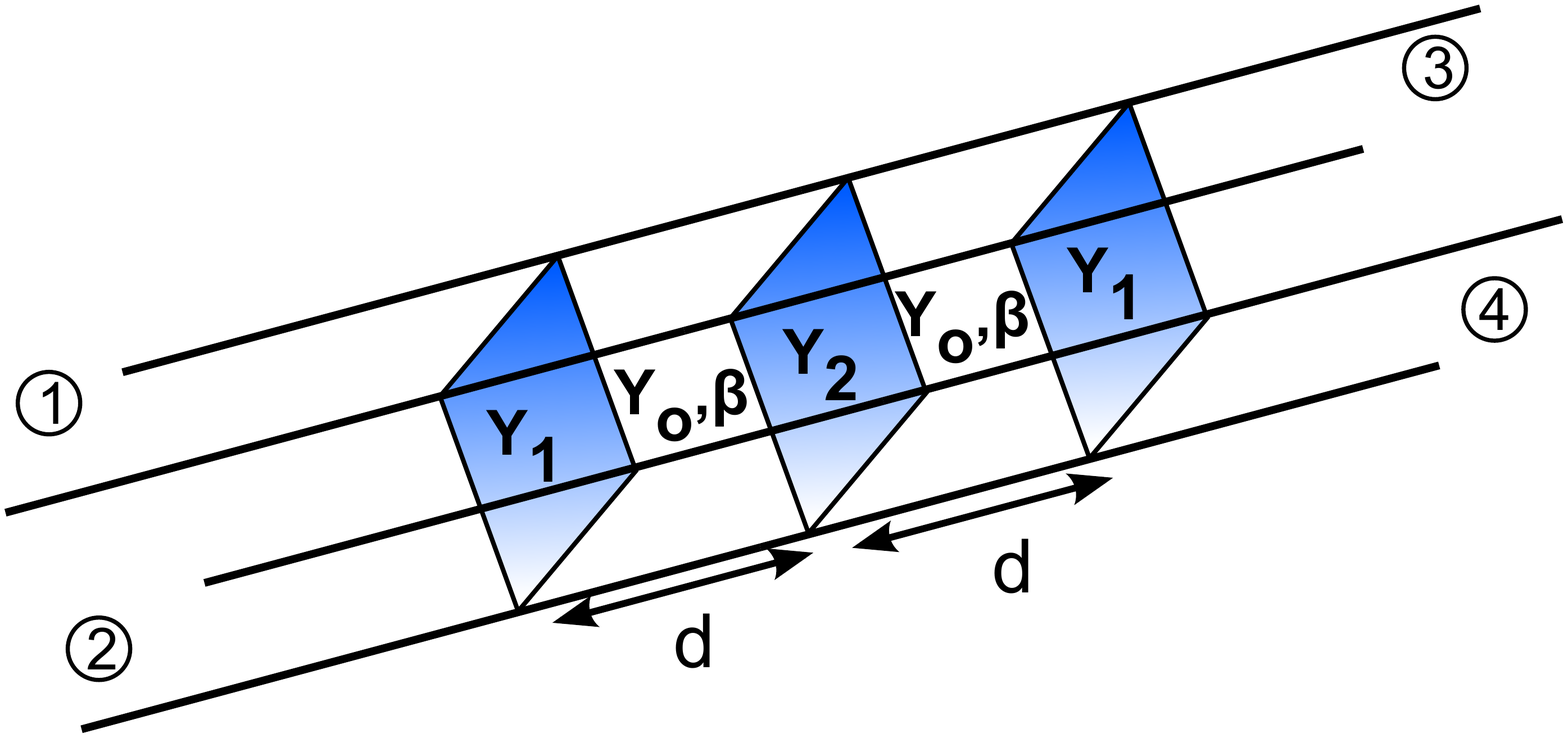}
	\label{fig:TensorTxArrayN=3}
}
\\
\subfloat[]
{
	\includegraphics[clip=true, trim=  0cm 3cm 0cm 3.5cm, scale=0.22]{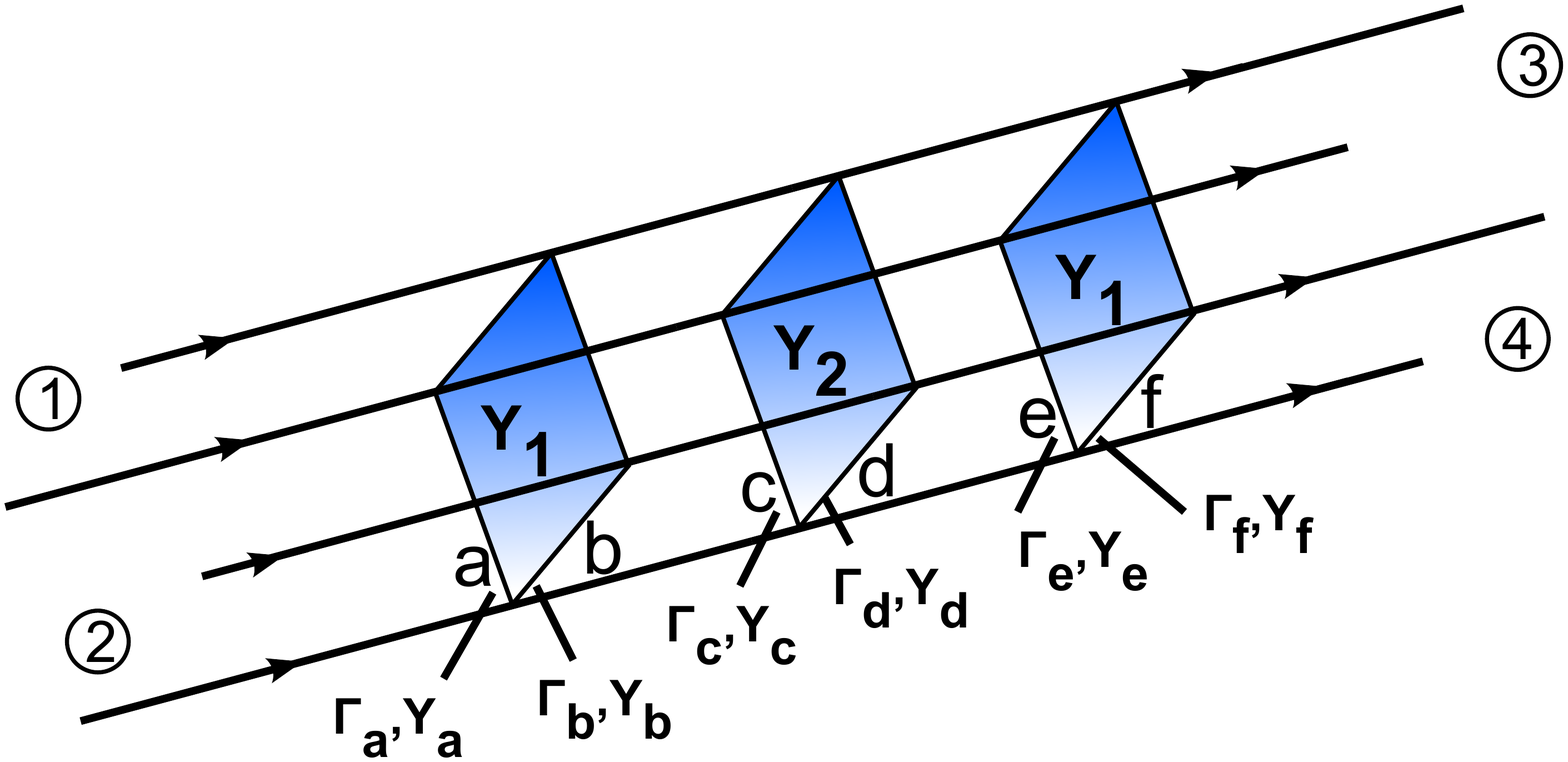}
	\label{fig:TensorTxArrayModelGammaN=3}
}
\caption{\protect\subref{fig:TensorTxArrayN=3} An equivalent circuit model for a symmetric tensor admittance transmitarray.  \protect\subref{fig:TensorTxArrayModelGammaN=3} The input admittance and reflection coefficient tensor defined at six different points within the symmetric tensor admittance transmitarray circuit model.}
\label{fig:TensorTxArray}
\end{figure}

An $N=3$ symmetric tensor impedance transmitarray is pictured in Fig.~\ref{fig:TensorTxArrayN=3}.  Using the two basic MTL operations described in the previous section we can formulate designs which have a reflection coefficient matrix of $\mathbf{0}$. Assuming that the dielectric constant and length $d$ of the dielectric spacers are fixed (usually due to fabrication constraints) we have two tensors to find with $\mathbf{Y_1}=\mathbf{Y_3}$ and $\mathbf{Y_2}$ each with three degrees of freedom.

To find the possible solutions of $\mathbf{Y_1}$ and $\mathbf{Y_2}$ we assume a value for $\mathbf{Y_1}$ and find the corresponding value of $\mathbf{Y_2}$ which satisfies the reflectionless property. Looking at Fig.~\ref{fig:TensorTxArrayModelGammaN=3}, we can see that at points $a$ and $f$ in the transmitarray, the reflection coefficient matrix is $\mathbf{0}$ and the input admittance matrix is given by \eqref{eq:CharImpMtx}. This is always true at $f$ since we are looking into a free-space MTL and true at $a$ as this is our desired result. From here we apply our two basic MTL operations to determine the input admittance at points $d$ and $c$ respectively. We do this by using the admittance of $\mathbf{Y_{1}}$ to find the input admittance matrix at points $b$ and $e$ using \eqref{eq:Yi}. We then translate this admittance along the MTL dielectric spacer to points $d$ and $c$ using \eqref{eq:YinTL}.  Once we have the admittance at $d$ and $c$ we can find the value of $\mathbf{Y_2}$ which is given to be
\begin{align}
\mathbf{Y_2}=\mathbf{Y_d}-\mathbf{Y_c}.
\end{align}
Once we have a solution of $\mathbf{Y_1}$ and $\mathbf{Y_2}$, we can find the complete S-parameter matrix which gives us the transmission through the transmitarray. This is done using the transfer matrices as shown in Appendix A. We can also formulate an analytical expression for the S-parameter matrix by enforcing the symmetric, reflectionless and lossless properties of the transmitarray on the S-parameter matrix itself \cite{Pozar,Selvanayagam2014}. This gives the following S-parameter matrix
\begin{align}
\label{eq:Smatrix}
\bm{S}=e^{-j\xi} \left[\begin{smallmatrix} 0 & \cos\Psi e^{-j\zeta} & 0 & \sin\Psi e^{-j\Phi} \\ \cos\Psi e^{-j\zeta}  & 0 &  \sin\Psi e^{-j\Phi} & 0\\0 &  \sin\Psi e^{-j\Phi} & 0 &  \cos\Psi\\\sin\Psi e^{-j\Phi}  & 0 &  \cos\Psi  & 0\end{smallmatrix}\right],
\end{align}
where $\xi$ is an arbitrary phase shift through the transmitarray. To ensure that the S-parameter matrix is unitary, we must have $\cos^2\Psi+\sin^2\Psi=1$ and $\zeta=2\Phi-(\pi+2n\pi), n \in \mathbb{Z}$.  This S-parameter matrix shows how a TE or TM polarization can be altered to any other polarization state.

To summarize, the procedure to solve for the admittance tensors $\mathbf{Y_{1}}$ and $\mathbf{Y_{2}}$ is given to be
\begin{enumerate}
\item Choose a value for $\mathbf{Y_1}$ via its eigenvalues $Y_{y,1}$ and $Y_{z,1}$ and its rotation angle $\gamma_1$
\item With this value, take the reflection coefficient and input admittances at points $a$ and $f$ and find their corresponding values at points $b$ and $e$ using the surface admittance of $\mathbf{Y_1}$ and \eqref{eq:Yi}.
\item The reflection coefficient and input admittances at $b$ and $e$ can be propagated along the transmission line interconnecting the transmitarray to points $c$ and $d$ using \eqref{eq:YinTL}. 
\item  Using the values of the admittance at points $c$ and $d$, the value of $\mathbf{Y_2}$ can be found which, for the given value of $\mathbf{Y_1}$, gives a transmitarray with zero reflections.
\item With the transmitarray specified, including all three admittance sheets, and the dielectric spacer, we can now determine the transmission coefficients through the transmitarray by finding the S-parameters as given in Appendix A. If the desired transmission parameters are not achieved we simply repeat the previous steps until they are.
\end{enumerate}

\begin{figure*}[!t]
\centering
\subfloat[]
{
	\includegraphics[clip=true, trim= 0cm 0cm 0cm 2cm, scale=0.25]{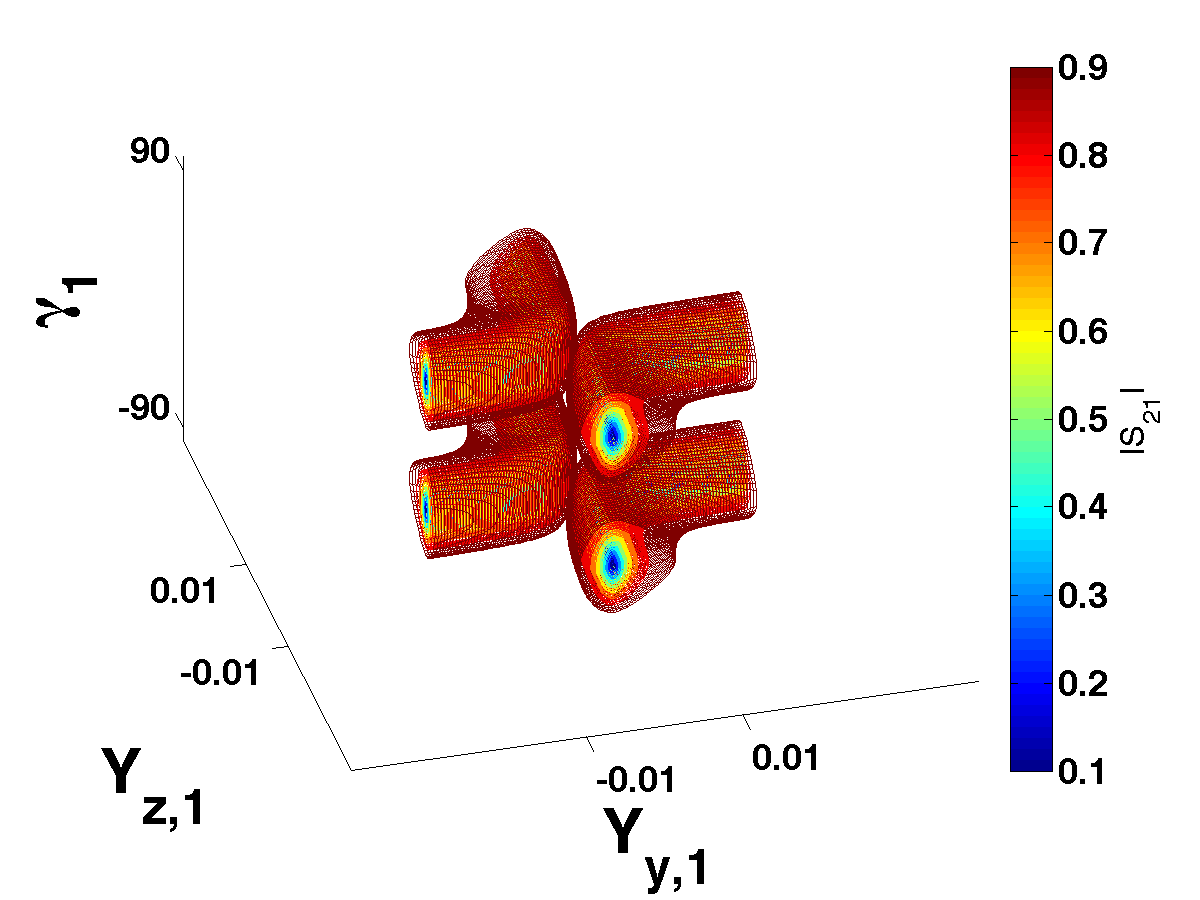}
	\label{fig:absS21vsY1}
}
\subfloat[]
{
	\includegraphics[clip=true, trim=  0cm 0cm 0cm 2cm, scale=0.25]{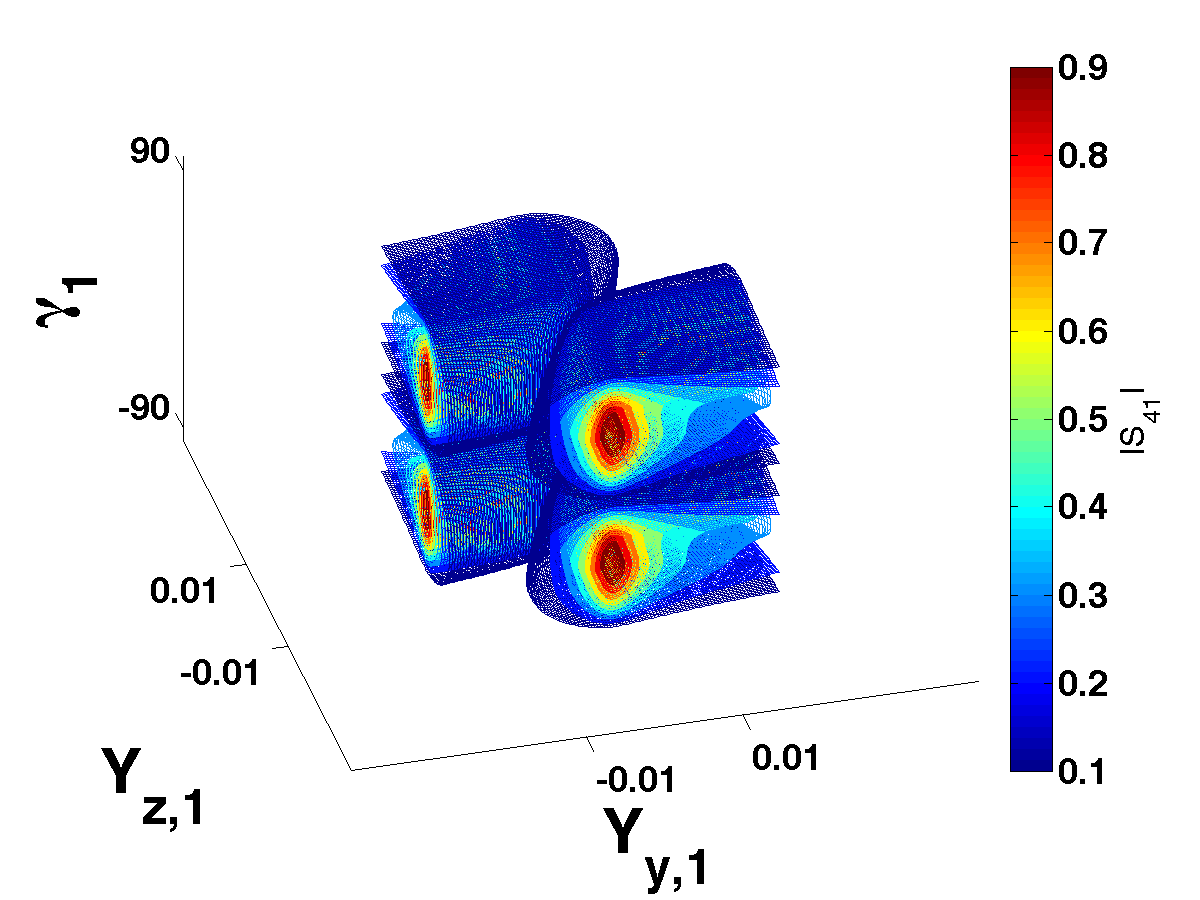}
	\label{fig:absS41vsY1}
}
\subfloat[]
{
	\includegraphics[clip=true, trim=  0cm 00cm 0cm 2cm, scale=0.25]{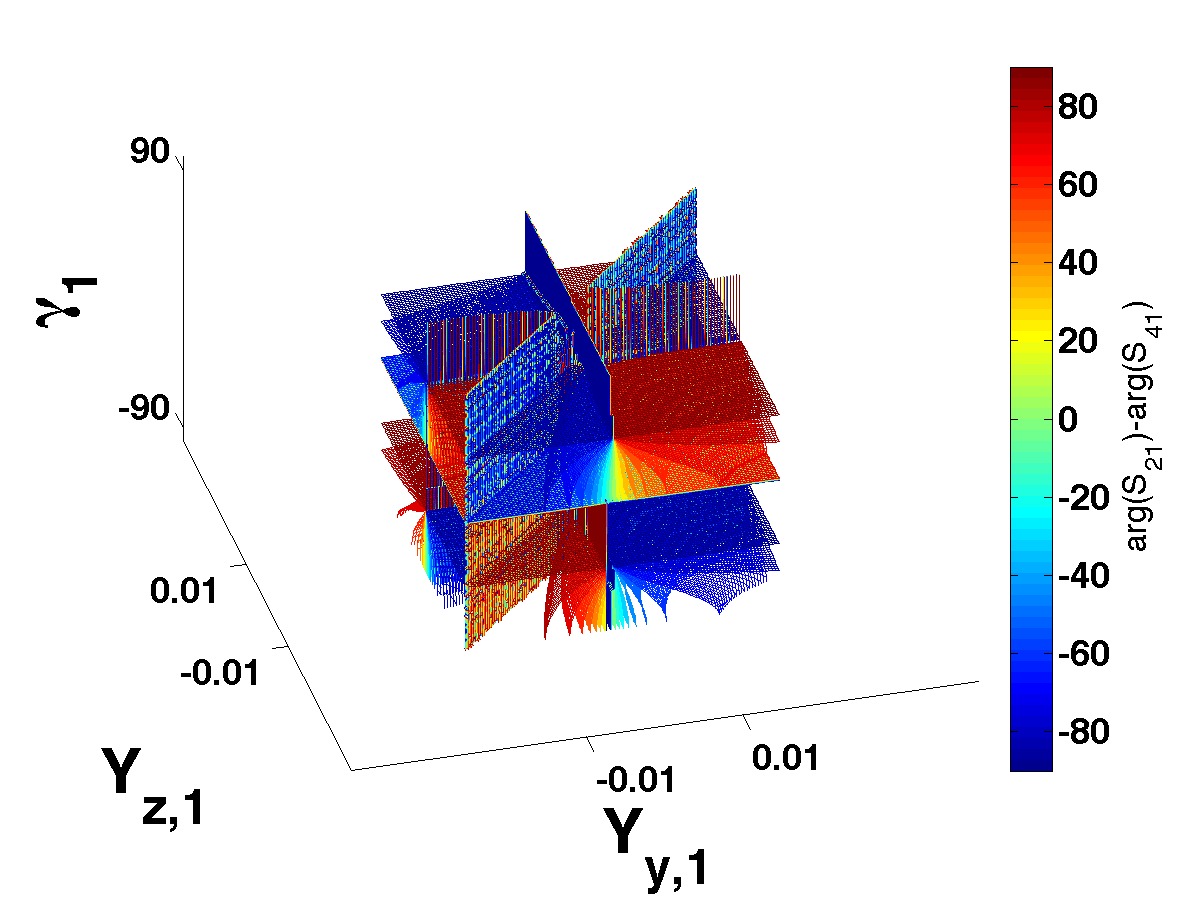}
	\label{fig:PhasevsY1}
}
\caption{The transmission parameters through the tensor impedance transmitarray as isosurfaces vs. $\mathbf{Y_1}$.  \protect\subref{fig:absS21vsY1}  $|S_{21}|$ \protect\subref{fig:absS41vsY1} $|S_{21}|$ \protect\subref{fig:PhasevsY1} $\arg(S_{21})-\arg(S_{41})$}
\label{fig:SvsY1}
\end{figure*}
Since we carry out this procedure as a function of $\mathbf{Y_1}$ we define a three dimensional space defined by the eigenvalues $Y_{y,1}$, $Y_{z,1}$ and rotation angle $\gamma_1$. We can then plot the calculated transmission parameters as a function of these variables in this three-dimensional space. This is shown in Fig.~\ref{fig:SvsY1} which maps out the entire solution space given by the S-parameter matrix in \eqref{eq:Smatrix}.

\subsection{Comments}
We make a couple of notes here about the solutions found for this symmetric tensor impedance transmitarray.  \begin{enumerate}
\item First we note that since $\mathbf{Y_3}=\mathbf{Y_1}$ the rotation angle $\gamma_1$ is the same for all three surface admittance layers.  Of the two MTL operations defined in Section~\ref{sec:Theory} only adding a shunt tensor admittance can change the rotation angle of the input admittance in the transmitarray. If $\mathbf{Y_1}$ and $\mathbf{Y_3}$ have the same rotation angle $\gamma_1$ then when translating the input admittance to points $d$ and $c$ we have the same rotation angle $\gamma_1$ and thus the rotation angle for $\mathbf{Y_2}$ is $\gamma_2=\gamma_1$. An intuitive physical interpretation then of the physical structure is of three layers of crossed-dipoles all rotated by the same angle.  We will see in the next section how this assumption can be broken for asymmetric tensor impedance transmitarrays. 

\item As stated above, the S-parameter matrix in \eqref{eq:Smatrix} shows how this symmetric transmitarray architecture can map a TE or TM polarization to any other desired state. This can be seen by the fact that the relative amplitude and phase of the transmitted field can be controlled based on the angles $\Psi$ and $\Phi$ which describe the polarization states on the Poincare sphere. Thus, this transmitarray can be designed then to alter the polarization state of a specific input polarization state to a desired output polarization state. For example, mapping a TE polarized field  to an elliptical polarization state. This is a generalization of the transmitarrays discussed in \cite{Joyal_2012,Pfeiffer_2013,He_2013} which are designed only for a linear polarization slanted at $45^{\circ}$. A similar result is also possible using tensor Huygens surfaces \cite{Selvanayagam2014}. However as we will see in Section~\ref{sec:AsymTxArray}, asymmetric transmitarrays can implement chiral effects which go beyond what is capable with a symmetric transmitarray.

\item Finally we note that because we have the ability to map a given input polarization state to any possible polarization state at the output we can design screens which implement an inhomogeneous spatial polarization variation such as orbital angular momentum. These inohomogeneous polarization states can be characterized by a higher order Poincare sphere \cite{Milione2011}. Metasurfaces such as \cite{Yu_etal_2011} implement these kind of inhomogeneous polarization states but suffer from large reflections and spurious diffraction orders. Here the symmetric tensor impedance transmitarray can generate any of the higher order states in \cite{Milione2011} for a specific input polarization state by locally designing the tensor admittances in the transmitarray without creating reflections or other diffraction orders.

\end{enumerate}

\subsection{Example - A Radial Polarization Screen}

\begin{figure}[!t]
\centering
\includegraphics[clip=true, trim= 0cm 0cm 0cm 0cm, scale=0.25]{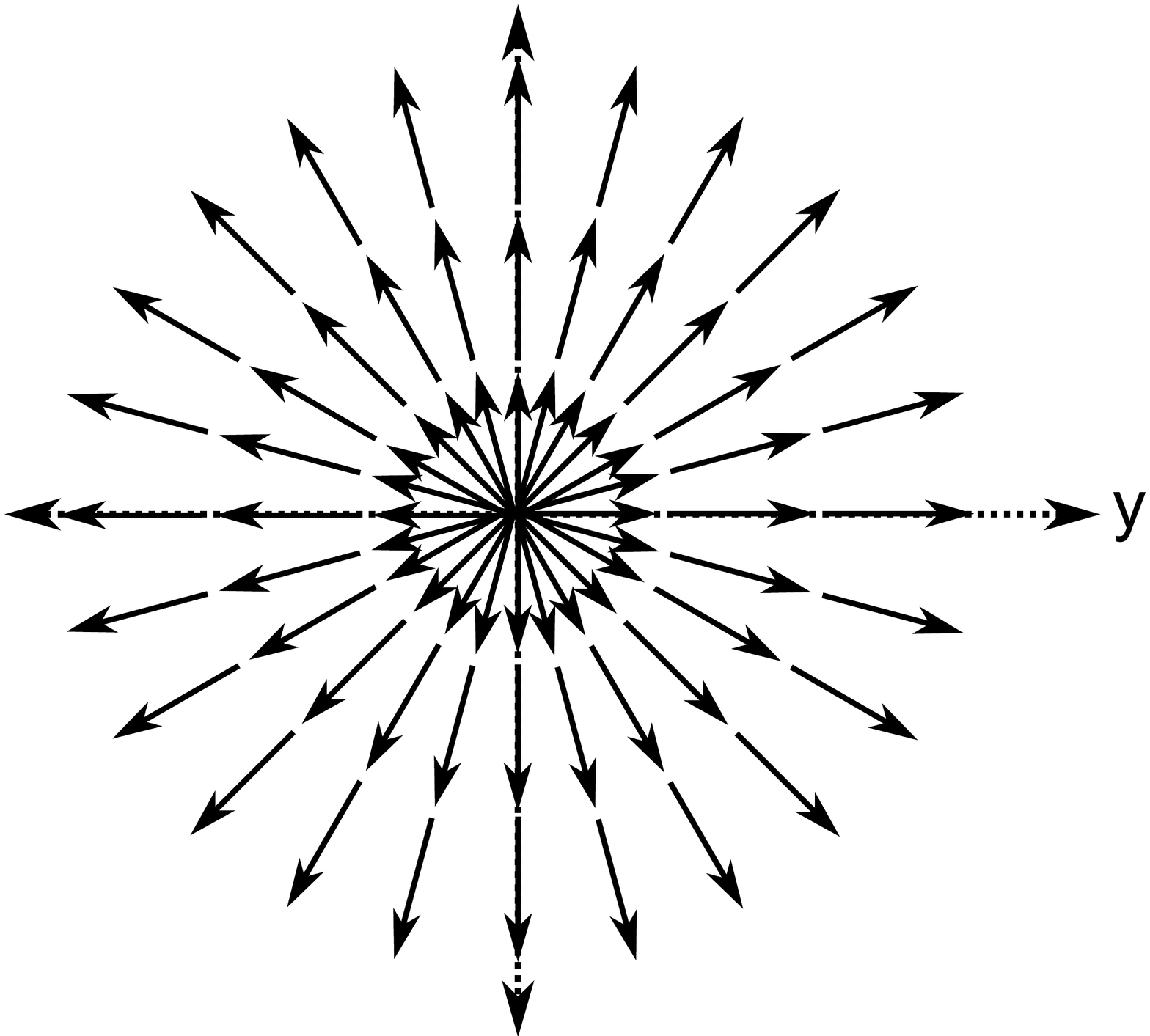}
\caption{The polarizaton of the electric field in a radially polarized beam.}
\label{fig:RadialPolSchem}
\end{figure}

As an example of the design procedure given in the above section as well as the polarization mapping capabilities of the symmetric tensor impedance transmitarray, we will design an $N=3$ symmetric tensor impedance transmitarray which takes a TE-polarized field and maps it into a radially polarized beam described in \cite{Milione2011}. A radially polarized beam consists of a beam whose electric field points in a radial direction on a plane transverse to the direction of propagation as illustrated in Fig.~\ref{fig:RadialPolSchem}. 

\begin{figure*}[!t]
\centering
\subfloat[]
{
	\includegraphics[clip=true, trim= 0cm 0cm 0cm 0cm, scale=0.22]{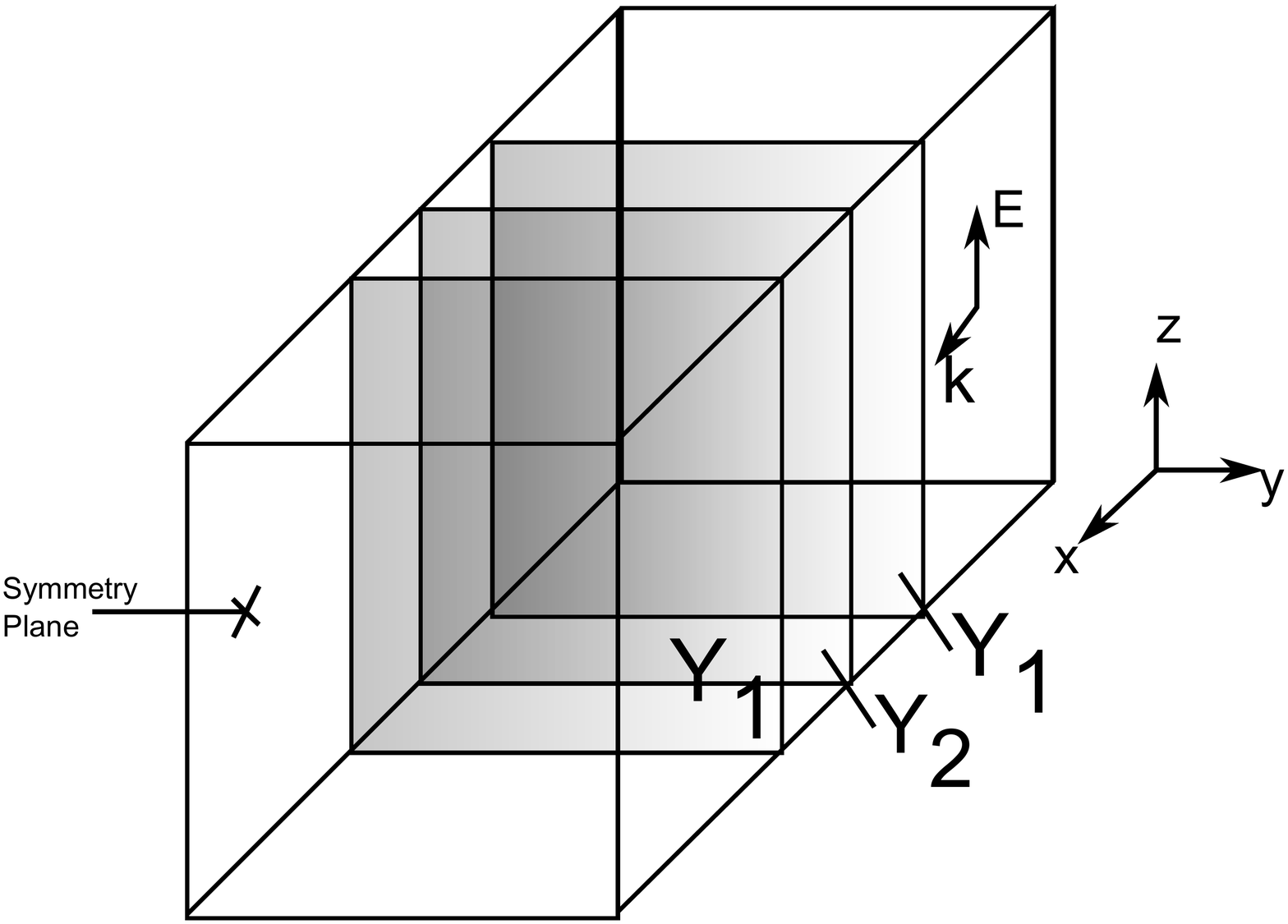}
	\label{fig:HFSSRadPolGeom}
}
\subfloat[]
{
	\includegraphics[clip=true, trim=  0cm 3cm 0cm 0cm, scale=0.15]{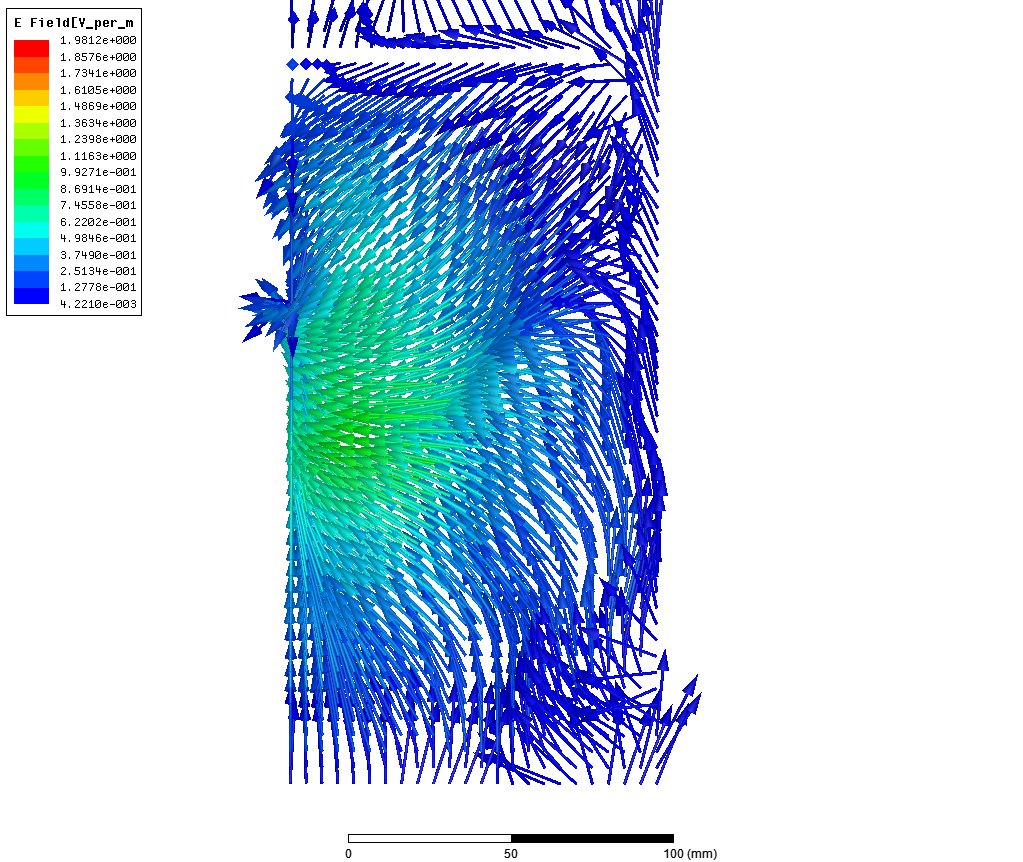}
	\label{fig:RadialPolTransverse}
}
\subfloat[]
{
	\includegraphics[clip=true, trim=  0cm 3cm 0cm 0cm, scale=0.15]{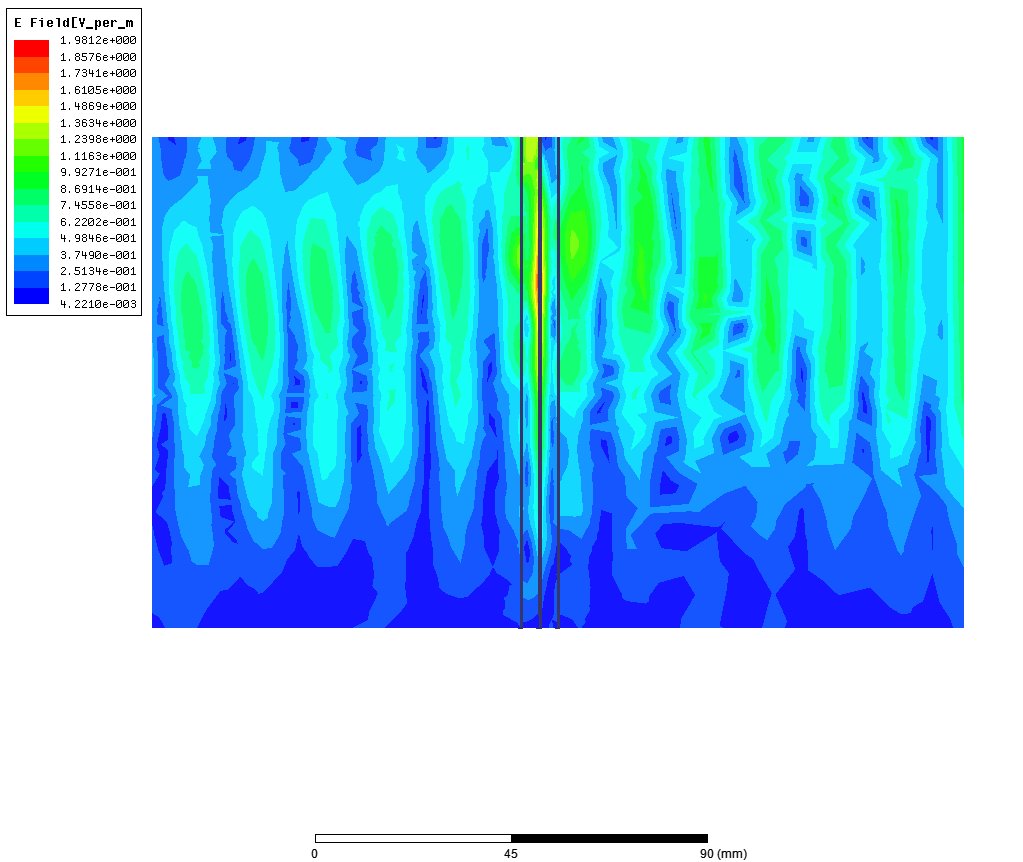}
	\label{fig:RadialPolPropagation}
}
\caption{\protect\subref{fig:HFSSRadPolGeom} The HFSS model All other boundaries of the computational domain besides the symmetry plane are terminated with radiation boundaries. \protect\subref{fig:RadialPolTransverse} The vectorial electric field on a transverse plane of the transmitted beam. We note the radial polarization of the beam at the center of the beam (ignoring the edge effects). \protect\subref{fig:RadialPolPropagation} The magnitude of the real part of the electric field vector behind and in front of the screen. We note that the input Gaussian beam is relatively undisturbed due to the minimal reflections. }
\label{fig:HFSSRadialPol}
\end{figure*}

We carry out this simulation using Ansys HFSS. Our simulation takes place at 10~GHz and our dielectric spacers are chosen to be $\lambda/7$ thick with a dielectric constant equal to free space. All of these choices are completely arbitrary and can be altered as necessary. The layers of the transmitarary sit parallel to  the $yz$-plane at $x={0,\lambda/7,2\lambda/7}$.  The domain is illuminated with a Gauussian beam due to the finite nature of the simulation with a waist of $2.5\lambda$ and electric field polarized along the $z$-axis (TE).  Each unit cell of the screen is $\lambda/5\times\lambda/5$.  To reduce the computational burden of the simulation we use symmetry boundaries to reduce the computational domain by half.  All of this is pictured in Fig.~\ref{fig:HFSSRadPolGeom}.

To find the required surface admittances for $\mathbf{Y_1}$ and $\mathbf{Y_2}$ we explore the solution space using the procedure given above and look for solutions which rotate a TE-polarized plane wave from $[0^{\circ},90^{\circ}]$ with the same absolute phase. This corresponds to an S-parameter matrix given by
\begin{align}
S=e^{-j\xi} \left[\begin{smallmatrix} 0 & \cos\Psi & 0 & \sin\Psi  \\ \cos\Psi   & 0 &  \sin\Psi & 0\\0 &  \sin\Psi  & 0 &  \cos\Psi\\\sin\Psi   & 0 &  \cos\Psi  & 0\end{smallmatrix}\right]
\end{align}
where $\Psi=\tan^{-1}(z/y)$ and $\Phi=0$ with $\xi$ some arbitrary but constant phase shift for the entire screen. Using our MTL model to determine the specific transmission properties we find $\mathbf{Y_1}$ and $\mathbf{Y_2}$ for each unit cell on the transmitarray. By using the anisotropic impedance boundary condition in HFSS we can create an ideal boundary for the surface admittance of each unit cell of each layer \cite{Quarfoth_2013}. 

Combining this all together and simulating, we get the field plot shown in Fig.~\ref{fig:RadialPolPropagation} and Fig.~\ref{fig:RadialPolTransverse}. Here we make some observations about our design.  Looking at the fields behind the screen in Fig.~\ref{fig:RadialPolPropagation}, we can see minimal reflections as the input Gaussian beam is relatively undisturbed. This validates the design procedure given above. When plotting the electric field on a plane transverse to the direction of propagation on the output side in Fig.~\ref{fig:RadialPolTransverse} we can see the radial polarization of the beam showing that this symmetric transmitarray does indeed provide the polarization control as intended. There is some fluctuation in the transmitted field amplitudes due to the coarse discretization of the screen. However this was unavoidable due to the computational size of the domain.

\section{Asymmetric  Tensor Impedance Transmitarrays}
\label{sec:AsymTxArray}

\begin{figure}[!t]
\centering
\subfloat[]
{
	\includegraphics[clip=true, trim= 0cm 4cm 0cm 3.5cm, scale=0.22]{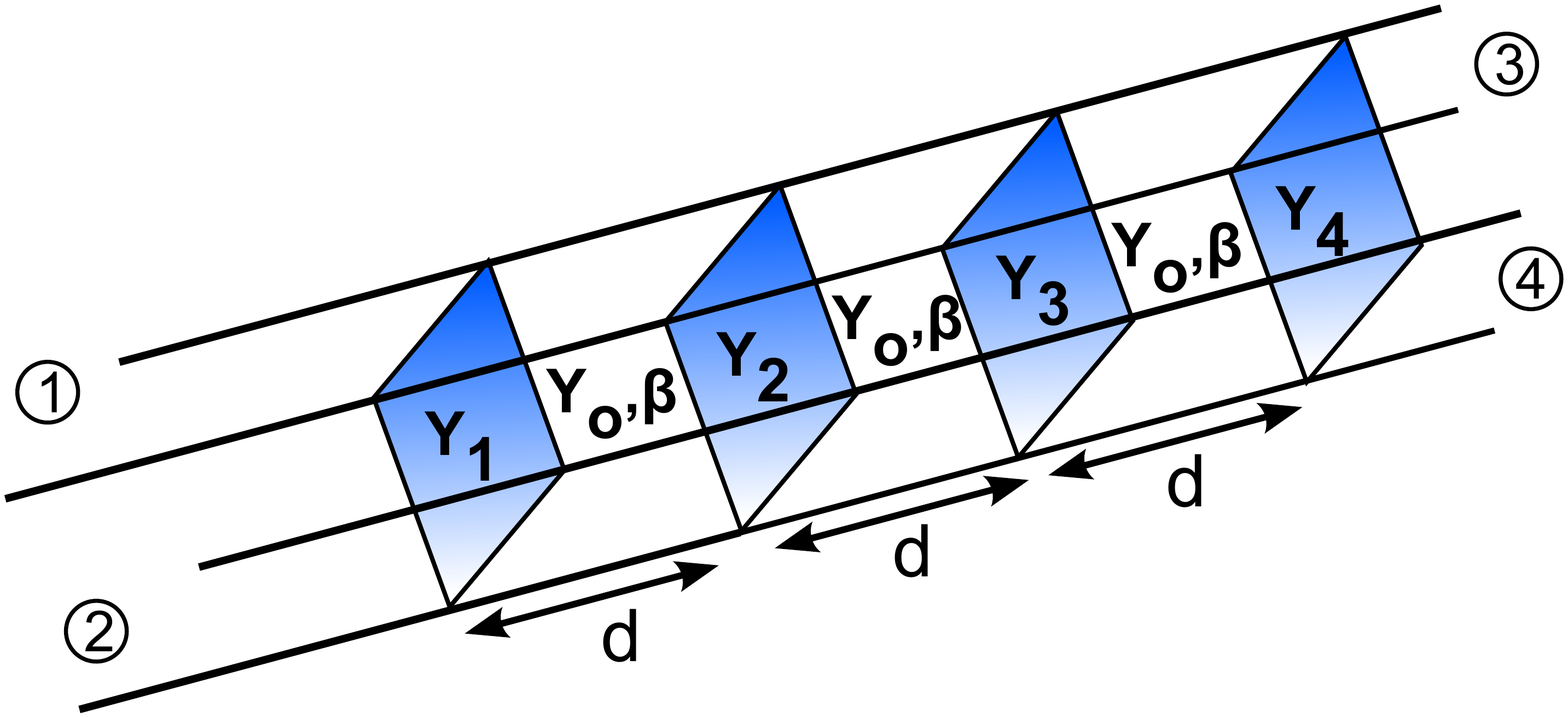}
	\label{fig:TensorTxArrayN=4}
}
\\
\subfloat[]
{
	\includegraphics[clip=true, trim=  0cm 3cm 0cm 3.5cm, scale=0.22]{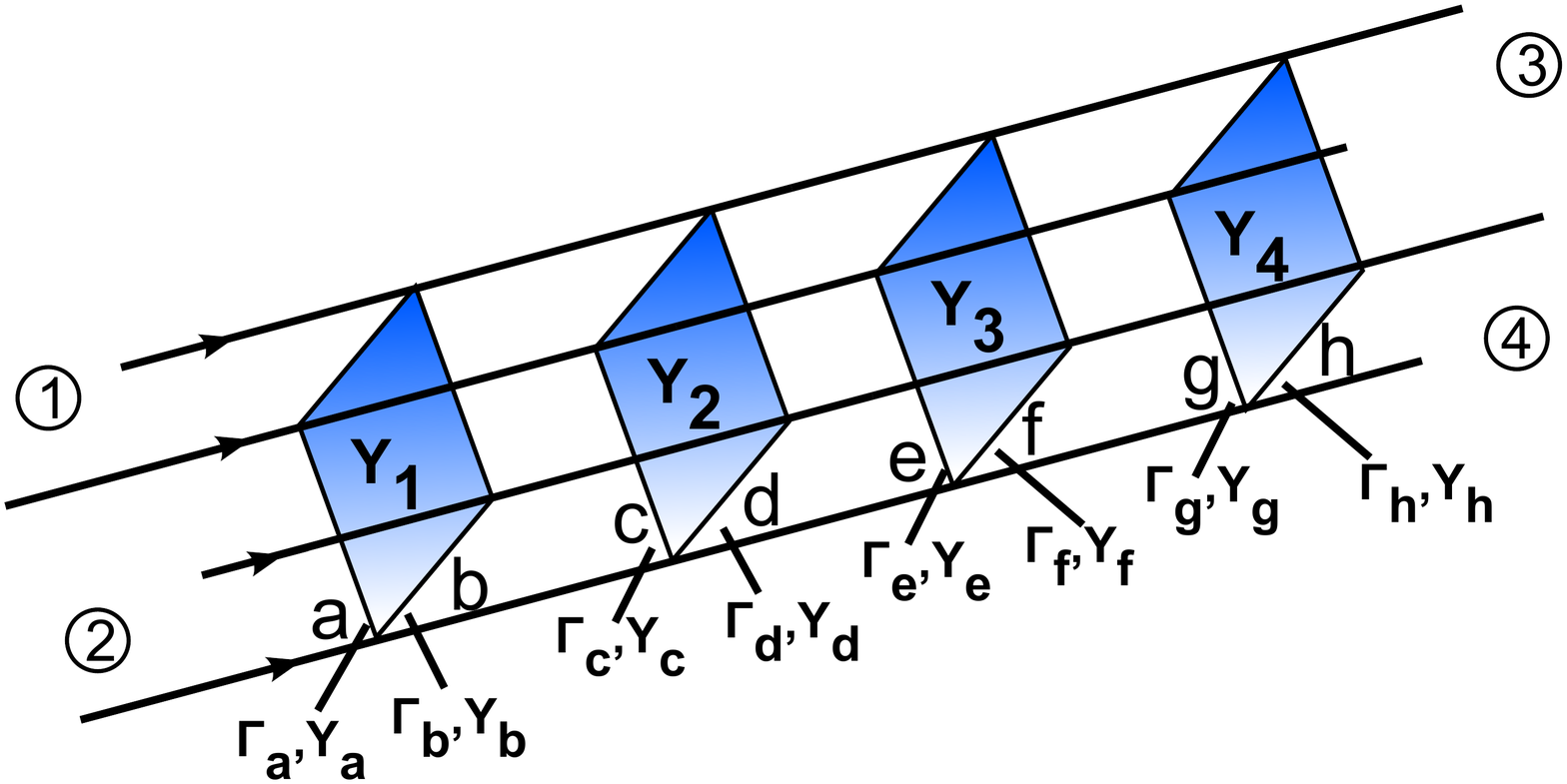}
	\label{fig:TensorTxArrayModelGammaN=4}
}
\caption{\protect\subref{fig:TensorTxArrayN=4} An equivalent circuit model for an asymmetric tensor admittance transmitarray.  \protect\subref{fig:TensorTxArrayModelGammaN=4} The input admittance and reflection coefficient tensor defined at eight different points within the asymmetric tensor admittance transmitarray circuit model.}
\label{fig:TensorTxArrayAsym}
\end{figure}

An $N=4$ asymmetric transmitarray is pictured in Fig.~\ref{fig:TensorTxArrayN=4}. Compared to the symmetric designs of the previous section, to design a reflectionless surface we need one more layer.  This makes sense as by making $\mathbf{Y_1}\neq\mathbf{Y_3}$ we no longer have $\mathbf{\Gamma}=\mathbf{0}$. Thus we need at least another layer, $\mathbf{Y_4}$ to compensate for this. 

However, if we already can design reflectionless tensor impedance transmitarray's using a symmetric design why do we need to investigate asymmetric designs? As mentioned in Section~\ref{sec:intro} in \cite{Zhao_etal_2011,Zhao_2012} it was shown how a stack of identical impedance surfaces each successively rotated by the same angle can mimic chiral behaviour which in \cite{Zhao_2012} was used to demonstrate circular polarization selectivity. This successive rotation of each layer causes each layer to have a different surface admittance and is thus asymmetric.  This demonstrated that chiral behaviour can be implemented and anlayzed by simply using surface admittance concepts only (However as stated above this approach only examined the effect of the rotation and not all the degrees of freedom of the admittance tensor ). Thus if we can analyze asymmetric tensor impedance transmitarrays and show how it can be designed to be reflectionless we can design these tensor impedance surfaces to implement chiral behaviour.

To see how this can be done for an $N=4$ layer structure let us examine the transmitarray in Fig.~\ref{fig:TensorTxArrayN=4}. Here we have four surface admittances $\mathbf{Y_1}$,$\mathbf{Y_2}$,$\mathbf{Y_3}$ and $\mathbf{Y_4}$ along with three dielectric spacers which we assume to be identical with the same length $d$ and dielectric constant $\varepsilon$.  We will follow a conceptually similar procedure used in Section~\ref{sec:SymTxArray} by charting the reflection coefficient and input admittance matrices through the transmitarray. This is annotated in Fig.~\ref{fig:TensorTxArrayModelGammaN=4} where the reflection coefficient and input admittance matrices are shown at 8 different points in the transmitarray Compared to the previous section however, we will require different analytical techniques to determine the surface admittance tensors of all the layers. 

To seek out a reflectionless design let us assume a value for $\mathbf{Y_1}$ and $\mathbf{Y_4}$ as long as $\mathbf{Y_1}\neq\mathbf{Y_4}$. We will see that this is a reasonable assumption with the examples that follow. We then have to find solutions for $\mathbf{Y_2}$ and $\mathbf{Y_{3}}$ so that the reflectionless property is maintained.

Starting with the input admittance at points $h$ and $a$ we again know that $\mathbf{Y_a}$ and $\mathbf{Y_h}$ are equal to \eqref{eq:CharImpMtx} with a reflection coefficient of $\mathbf{0}$. We can then apply our two basic MTL operations to find the input admittance at points $c$ and $f$. For point $c$ this is done by adding the admittance of $\mathbf{Y_1}$ and translating the impedance along the MTL line representing the dielectric spacer to $c$. Likewise for point $f$. With the input admittance at points $c$ and $f$ the surface admittance of $\mathbf{Y_2}$ and $\mathbf{Y_3}$ and the middle dielectric spacer separate these two points.  

To find the value of $\mathbf{Y_2}$ and $\mathbf{Y_3}$ which will enforce a $\mathbf{0}$ reflection coefficient we can make a couple of helpful observations.  First we note that at points $d$ and $e$ the real part of the input admittance is the same as points $c$ and $f$ respectively.  This is because $\mathbf{Y_2}$ and $\mathbf{Y_3}$ only alter the imaginary part of the input admittance. We also note that translating the input admittance on the fixed dielectric spacer from point $e$ to $d$ must allow for the real part of the admittance at $e$ to translate to the real part of the admittance at $d$. Since translating an admittance along an MTL alters both the real and imaginary parts of the admittance, and since we are assuming that the dielectric spacer is fixed, this constraint gives us a way of evaluating the total input admittance at $e$. This is because for the real parts at points $e$ and $d$ to be consistent, the imaginary part of the admittance at $e$ must also take on a specific value when the input admittance is translated along the MTL. Enforcing this will allow for the imaginary part at point $e$ to be determined. Once we have that we can find $\mathbf{Y_3}$. We can then find the imaginary part of the admittance at $d$ and find $\mathbf{Y_2}$.  

To express this mathematically we can use \eqref{eq:YinTL}.  At point $e$ we have the input admittance given by $\mathbf{Y_e}=\mathbf{G_e}+j\mathbf{B_e}$. As stated we know $\mathbf{G_e}=\mathbf{G_f}$ but we would like to find $\mathbf{B_e}$.
We can set up the relationship between points $d$ and $e$ which is given to be
\begin{align}
\label{eq:YdTLYe}
\mathbf{G_d}+j \mathbf{B_d}=\frac{1}{\eta}\left( \mathbf{G_e}+j\mathbf{B_e} +j\mathbf{Y_o}\tan{\beta d}  \right) \\ \nonumber \left(  \mathbf{Y_o} +j(\mathbf{G_e}+j\mathbf{B_e})\tan(\beta d)    \right)^{-1},
\end{align}
where $\mathbf{Y_d}=\mathbf{G_d}+j \mathbf{B_d}$ and we know that $\mathbf{G_d}=\mathbf{G_c}$. Separating the above equation into real and imaginary parts and looking at the real part only we get the following equation as a function of $\mathbf{B_e}$
\begin{align}
\label{eq:MTLRiccati}
-\frac{1}{\eta}\mathbf{G_e}^{-1}\mathbf{B_e} -\frac{1}{\eta}\mathbf{B_e}\mathbf{G_e}^{-1} + &\frac{1}{\tan(\beta d)} \mathbf{B_e} \mathbf{G_e}^{-1} \mathbf{B_e} + \\  \nonumber  \frac{1}{\eta^2 \tan(\beta d)} \mathbf{G_e}^{-1}+\tan(\beta d)\mathbf{G_e} - \\ \nonumber \frac{1}{\eta}\mathbf{G_d}^{-1}\left( \frac{1}{\eta \tan(\beta d)}\mathbf{E}+ \frac{\tan(\beta d)}{\eta}\mathbf{E} \right) &= \mathbf{0}.
\end{align}
This equation is in the form of 
\begin{align}
\mathbf{0}=\mathbf{A}^{T}\mathbf{B_e} + \mathbf{B_e}\mathbf{A}-\mathbf{B_e}\mathbf{C}\mathbf{B_e}+\mathbf{Q}
\end{align}
where the coefficients $\mathbf{A}$, $\mathbf{C}$ and $\mathbf{Q}$ are given to be
\begin{align}
\mathbf{A}&=-\frac{1}{\eta}\mathbf{G_e}^{-1},\\
\mathbf{C}&=\frac{1}{\tan(\beta d)} \mathbf{G_e}^{-1},\\
\mathbf{Q}&= \frac{1}{\eta^2 \tan(\beta d)} \mathbf{G_e}^{-1}+\tan(\beta d)\mathbf{G_e} - \\ & \nonumber \frac{1}{\eta}\mathbf{G_d}^{-1}\left( \frac{1}{\eta \tan(\beta d)}\mathbf{E}+ \frac{\tan(\beta d)}{\eta}\mathbf{E} \right)
\end{align}
This equation is known to be the algebraic Riccati equation which has applications in control theory as well as MTL theory \cite{Potter1966,Assante2012}. Solutions to this equation can be constructed explicitly from the eigenvectors of a block matrix of the coefficients \cite{Potter1966} or through numerical tools \footnote{For example, in MATLAB this equation can be solved using the \texttt{care}  command}. Doing this gives us $\mathbf{B_e}$ the susceptance at $e$. We know now $\mathbf{Y_e}=\mathbf{G_e}+j\mathbf{B_e}$ and can find $\mathbf{Y_3}$ to be
\begin{align}
\label{eq:Y3}
\mathbf{Y_3}=\mathbf{Y_f}-\mathbf{Y_e}.
\end{align}
Then using \eqref{eq:YdTLYe} we can find $\mathbf{Y_d}$ and $\mathbf{Y_2}$ is then given to be
\begin{align}
\label{eq:Y2}
\mathbf{Y_2}=\mathbf{Y_d}-\mathbf{Y_c}.
\end{align}
This procedure gives us the four surface admittances which form an asymmetric transmitarray that is reflectionless. 

To find the transmission coefficients through this transmitarray we can use the same procedure in Appendix A which allows us to calculate the S-parameters using the transfer matrices of the four-layer transmitarray.

To summarize this, we can follow the following procedure to design a reflectionless asymmetric four-layer tensor admittance transmitarray
\begin{enumerate}
\item We assume a value for $\mathbf{Y_1}$ and $\mathbf{Y_4}$ via the rotation angles and eignevalues of each tensor.
\item Knowing the input admittance at $a$ and $h$ we can calculate the input admittance at $b$ and $g$ and subsequently $c$ and $f$ using our two basic MTL operations
\item The real part of the input admittance at $d$ and $e$ is known from points $c$ and $f$. The admittance at points $d$ and $e$ is also constrained due to the dielectric spacer between them. This is given by the algebraic Riccati equation in \eqref{eq:MTLRiccati} which we solve to find the imaginary part of the admittance at $e$.
\item Finally with the admittance at $e$ known, we can use \eqref{eq:Y3} to find $\mathbf{Y_3}$. We can then use the admittance at $e$ to determine the admittance at $d$ so that \eqref{eq:Y2} can be used to find $\mathbf{Y_2}$. This gives us the admittance for all four layers.
\item We can evaluate the transmission through the structure using the methods in Appendix A. If the desired transmission values are not achieved we then repeat the previous steps until they are.
\end{enumerate}

The S-parameters in general for such a surface are characterized by the following lossless matrix,
\begin{align}
\label{eq:SmatrixAsym}
\bm{S}=e^{-j\xi} \left[\begin{smallmatrix} 0 & \cos\Psi  & 0 & \sin\Psi e^{-j\Phi} \\ \cos\Psi  & 0 &  -\sin\Psi e^{j\Phi} & 0\\0 &  -\sin\Psi e^{j\Phi} & 0 &  \cos\Psi\\\sin\Psi e^{-j\Phi}  & 0 &  \cos\Psi  & 0\end{smallmatrix}\right],
\end{align}
This looks similar to \eqref{eq:Smatrix} and in fact, using this asymmetric transmitarray can allow us to achieve many of the same polarization controlling properties as a symmetric transmitarray. Thus we can alter a specific input polarization state into another polarization state. However in \eqref{eq:SmatrixAsym} we also note that the cross coupling transmission coefficients have a $\pi$ phase shift between them. This will allow for the implementation of chiral effects such as polarization rotator. (By relaxing the reflectionless requirement we can also implement a circular polarizer). The physical interpretation of this different cross coupling is due to the  admittance of each layer having a different rotation angle $\gamma$. 

\subsection{Examples}
To demonstrate the design procedure given above for asymmetric tensor impedance transmitarrays. We will look at two examples which implement chiral behaviour.  These examples are for achieving linear polarization rotation and circular polarization selectivity \cite{Niemi2013}.  We will demonstrate both here.

\begin{figure}[!t]
\centering
\includegraphics[clip=true, trim= 0cm 0cm 0cm 0cm, scale=0.25]{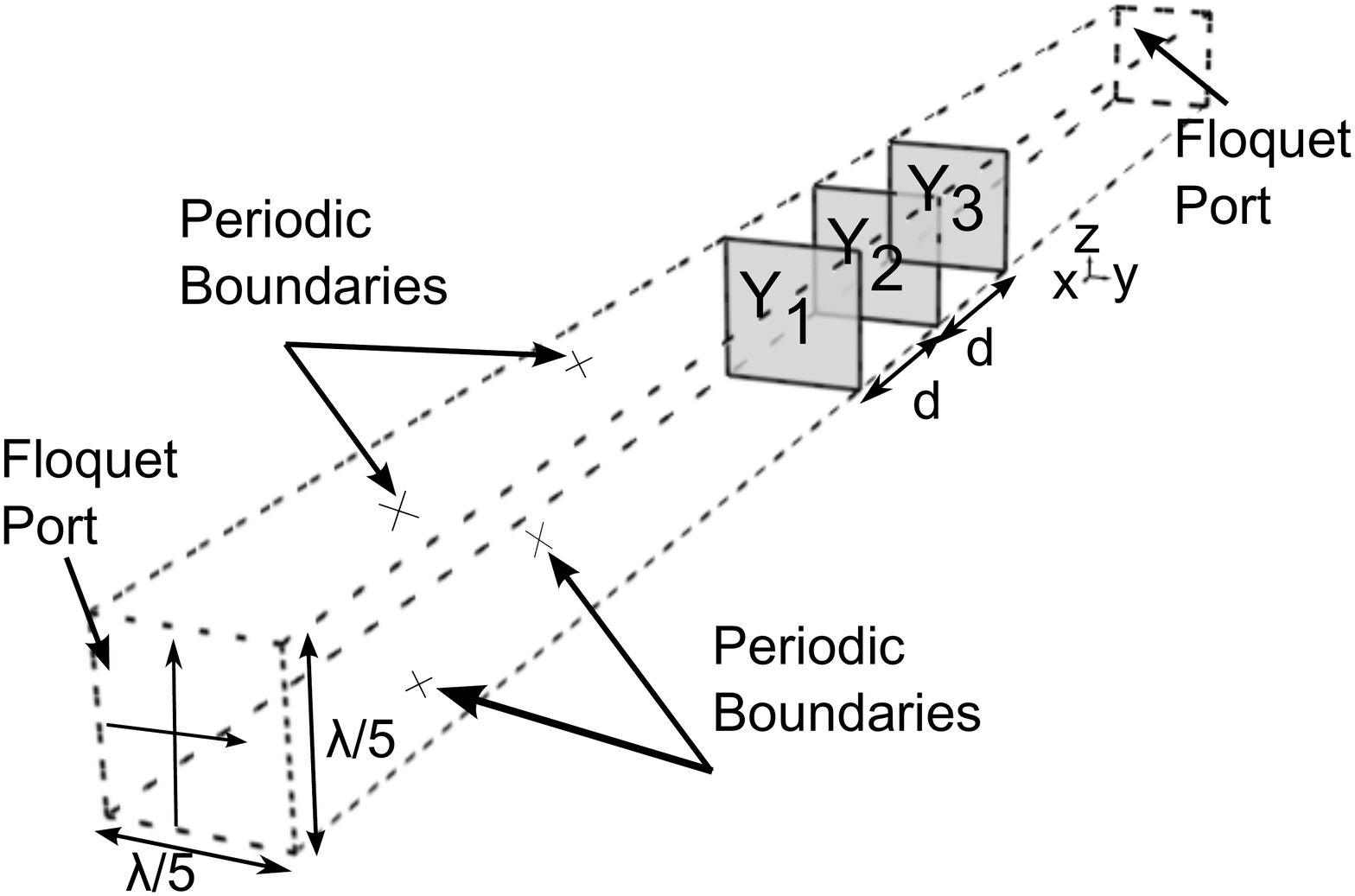}
\caption{The computational model for the $N=4$ layer polarization rotator and the $N=3$ circular polarizer. ($N=3$ model shown).}
\label{fig:HFSSgeom}
\end{figure}

As in the last section, all of our examples will take place at 10~GHz and are simulated in Ansys HFSS. Because these examples are dealing with homogeneous screens we can simply simulate one unit cell surrounded by periodic boundary conditions as shown in Fig.~\ref{fig:HFSSgeom}. The domain is then excited by periodic ports (referred to as Floquet ports in HFSS). Each unit cell is $\lambda/5 \times \lambda/5$. Again for each layer of the transmitarray we will use anisotropic impedance boundary conditions in HFSS which represents an ideal surface impedance.

\subsubsection{A Polarization Rotator}

\begin{figure*}[!t]
\centering
\subfloat[]
{
	\includegraphics[clip=true, trim= 0cm 5cm 0cm 4cm, scale=0.22]{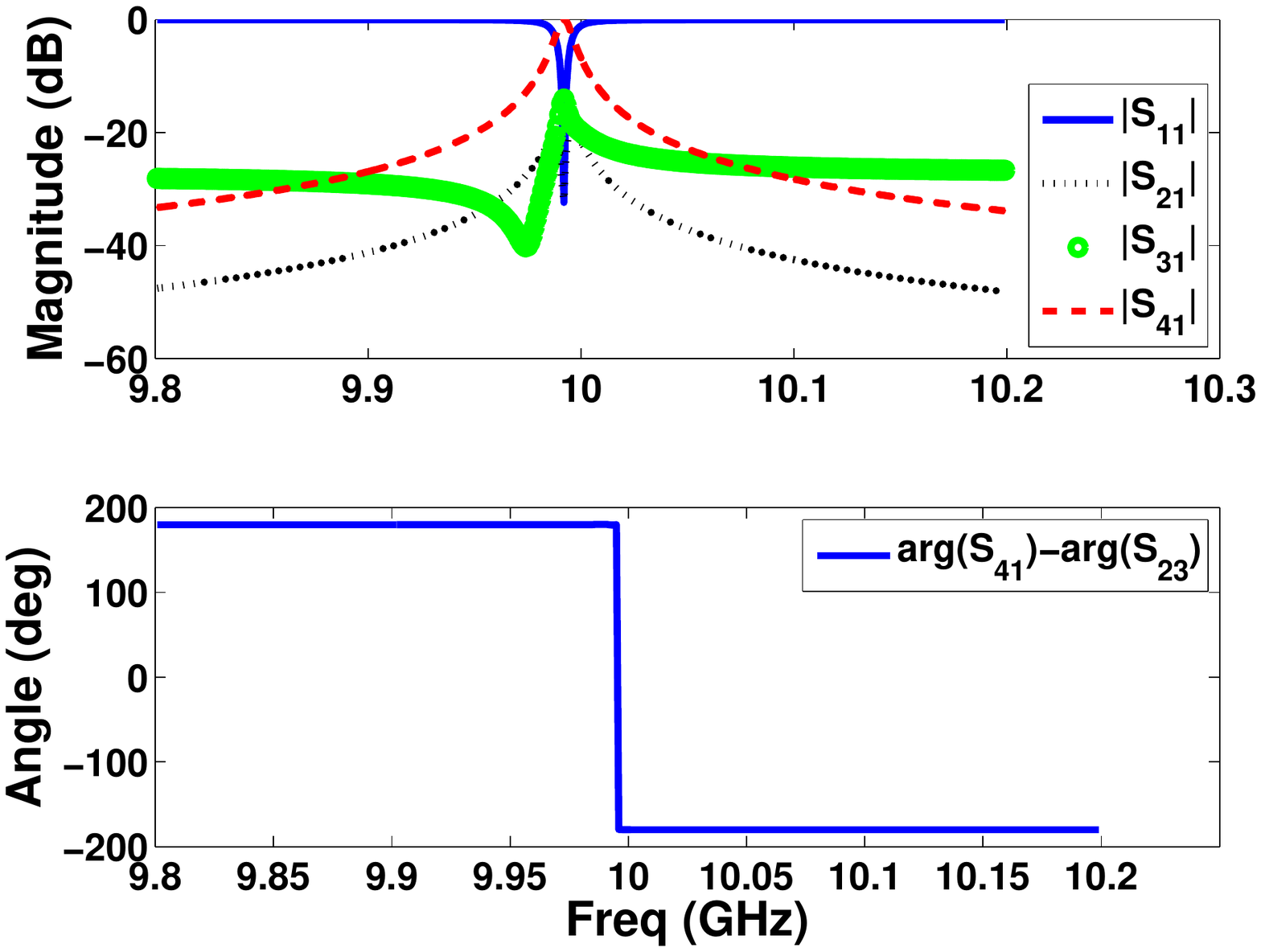}
	\label{fig:SparamTwistPolRotMag}
}
\subfloat[]
{
	\includegraphics[clip=true, trim=  0cm 2cm 0cm 0cm, scale=0.13]{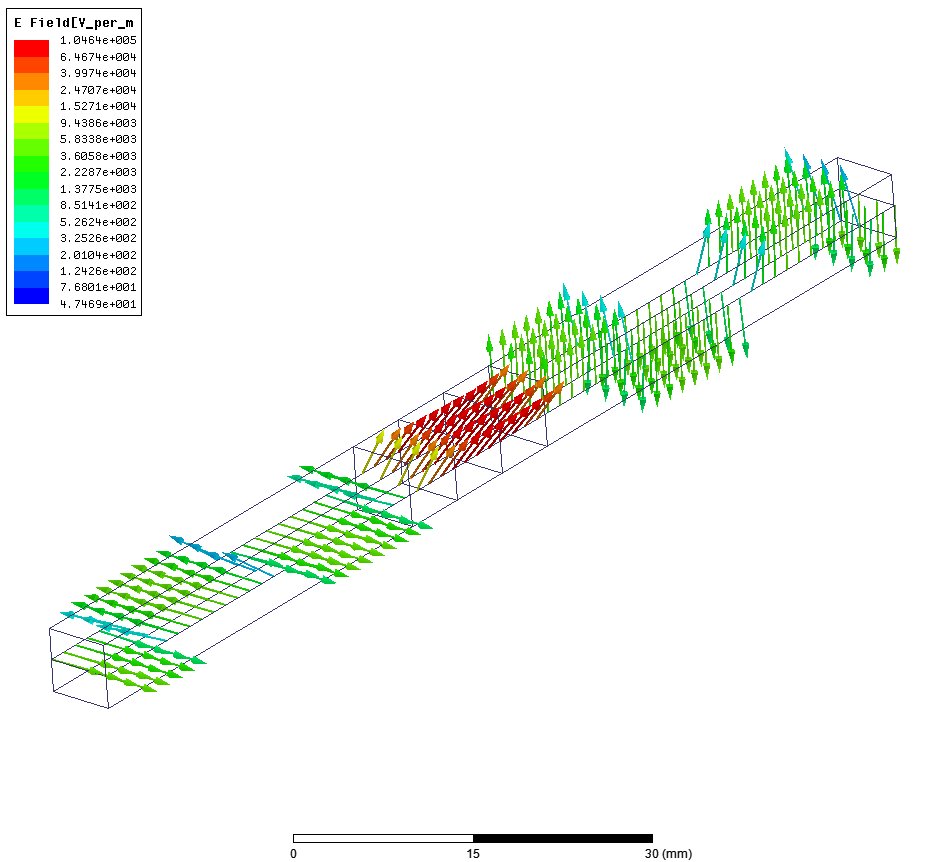}
	\label{fig:PolRotEfieldTE}
}
\subfloat[]
{
	\includegraphics[clip=true, trim=  0cm 2cm 0cm 0cm, scale=0.13]{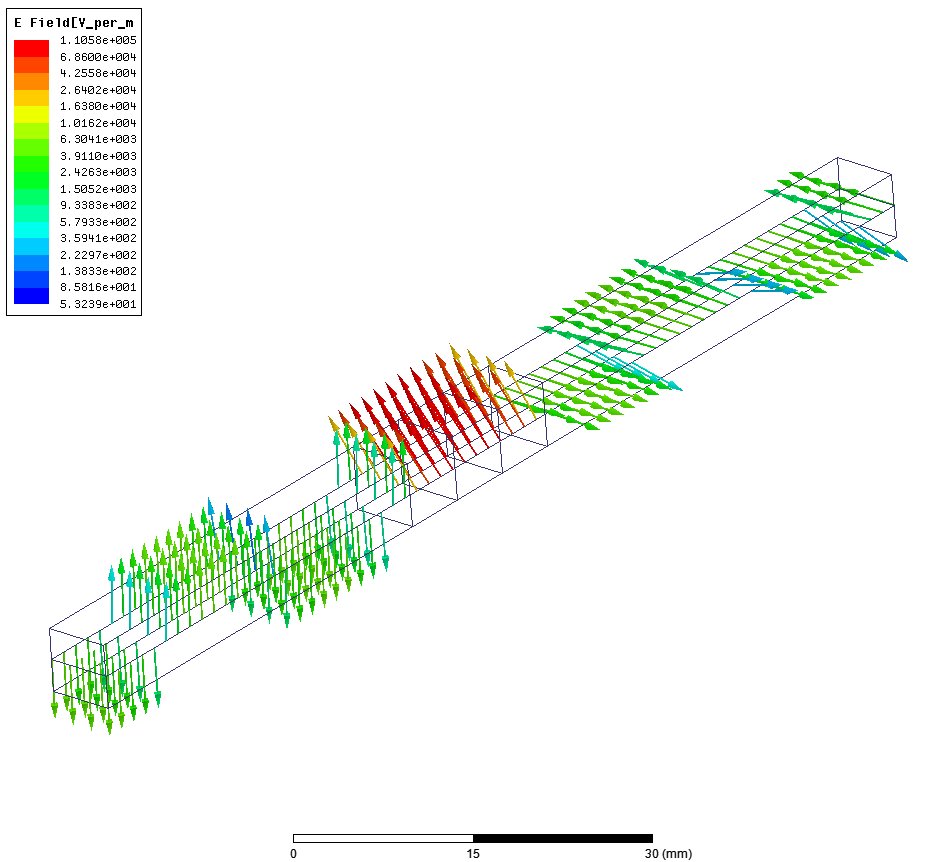}
	\label{fig:PolRotEfieldTM}
}
\subfloat[]
{
	\includegraphics[clip=true, trim=  0cm 2cm 0cm 0cm, scale=0.13]{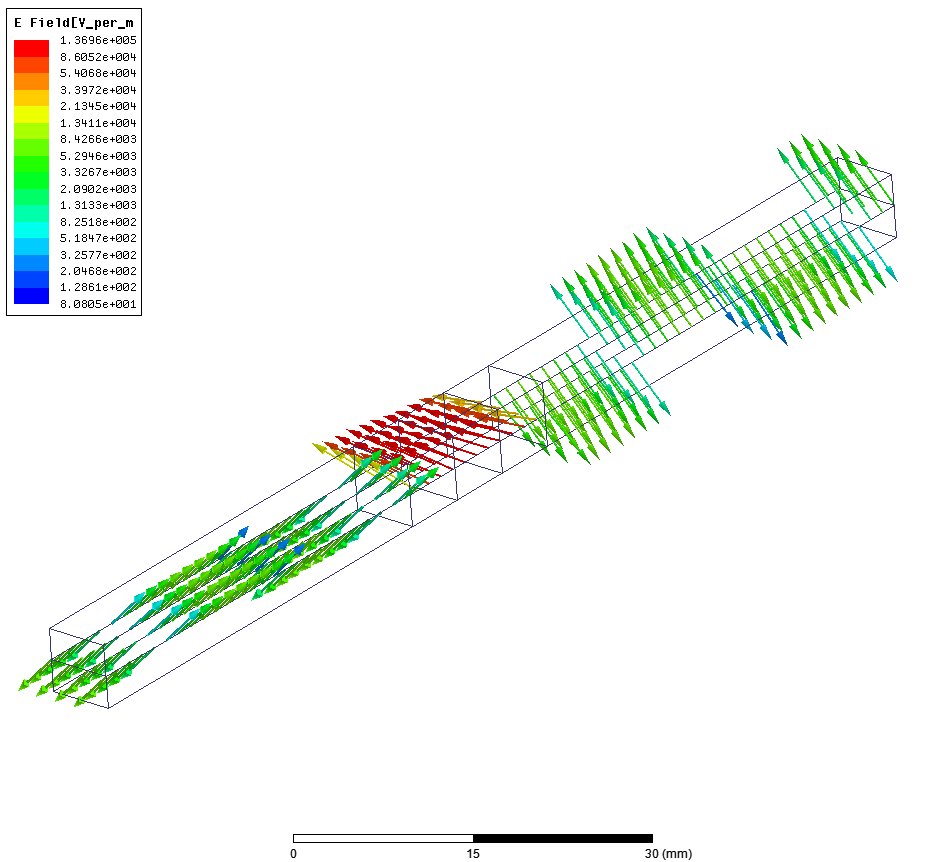}
	\label{fig:PolRotEfield45}
}
\caption{\protect\subref{fig:SparamTwistPolRotMag} The S-parameters of the polarization rotator. The design is narrowband due to our choice of $\mathbf{Y_1}$ and $\mathbf{y_4}$.  The cross-polarized transmission peaks at 10~GHz where the transmitarray is reflectionless. In the phase plot we can see the $\pi$ phase difference between the two cross-polarized transmission coefficients \protect\subref{fig:PolRotEfieldTE} The electric field in the unit cell when excited a TE-polarized wave. The transmitted field is a TM-polarized field ($90^{\circ}$ rotation)  \protect\subref{fig:PolRotEfieldTM} The electric field in the unit cell when excited by a TM-polarized field. \protect\subref{fig:PolRotEfield45} The electric field in the unit cell when excited by a $45^{\circ}$ slant polarization. It is also rotated by $90^{\circ}$. }
\label{fig:HFSSCPSS}
\end{figure*}

For a polarization rotator the S-parameters of the surface are given by
\begin{equation}
\label{eq:SparamPolRot}
S=e^{-j\xi} \left[ \begin{smallmatrix} 0 & \cos\Psi & 0 & \sin\Psi  \\ \cos\Psi   & 0 &  -\sin\Psi & 0 \\ 0 &  -\sin\Psi  & 0 &  \cos\Psi \\ \sin\Psi   & 0 &  \cos\Psi  & 0 \end{smallmatrix}\right]
\end{equation}
As stated, we note the $\pi$ phase shift between the cross-polarized transmission of the TE and TM fields. We  also note that this kind of transmitarray can take any linear polariztaion and rotate it by the same angle $\Psi$ whereas the example in Section~\ref{sec:SymTxArray} was designed for one specific input polarization. This is the added flexibility of implementing the chiral behaviour with the asymmetric transmitarray.

To implement a design of a surface which realizes this S-parameter matrix we follow the procedure given above. Here we set the spacing between the admittance surfaces to be $d=\lambda/5$.  Since we must assume a value for both $\mathbf{Y_1}$ and $\mathbf{Y_4}$ we make the simplest choice which breaks the symmetry of the transmitarray. We choose a value for $\mathbf{Y_1}$ based on its three degrees of freedom, its eigenvalues $Y_{1,y}$ and $Y_{1,z}$ and rotation angle $\gamma_1$. For $\mathbf{Y_4}$ we then choose the same eigenvalues $Y_{4,y}=Y_{1,y}$ and $Y_{4,z}=Y_{1,z}$ but $\gamma_4=-\gamma_1$.  The consequence of this is that we find highly resonant designs which are inherently narrowband. While this is not very practical for actual designs it is sufficient for our purposes here to demonstrate the concept. Further work will involve exploring suitable choices for $\mathbf{Y_1}$ and $\mathbf{Y_4}$ that lead to more optimal designs. With this choice of $\mathbf{Y_1}$ and $\mathbf{Y_4}$ we can then find $\mathbf{Y_2}$ and $\mathbf{Y_3}$. 

The example we choose to simulate is one which rotates the linear polarization by $\Psi=\pi/2$. One possible solution is given by surface admittances $\mathbf{Y_1}=\left[\begin{array}{cc} j5.16 \times 10^{-2} &-j6.44\times 10^{-2} \\ -j6.44\times 10^{-2}  & j7.29\times 10^{-2} \end{array} \right]\mho$, $\mathbf{Y_2}=\left[\begin{array}{cc} -j1.04 \times 10^{-3} & j4.51\times 10^{-3} \\ j4.51\times 10^{-3}  & -j1.16\times 10^{-3} \end{array} \right]\mho$, $\mathbf{Y_3}=\left[\begin{array}{cc} -j1.04 \times 10^{-3} & -j4.51\times 10^{-3} \\ -j4.51\times 10^{-3}  & -j1.16\times 10^{-3} \end{array} \right]\mho$, and $\mathbf{Y_4}=\left[\begin{array}{cc} j5.16 \times 10^{-2} & j6.44\times 10^{-2} \\ j6.44\times 10^{-2}  & j7.29\times 10^{-2} \end{array} \right]\mho$.  

We note that the rotation angle of each surface is given by $\gamma_1=-22.99^{\circ}$, $\gamma_2=41.13^{\circ}$, $\gamma_3=-41.13^{\circ}$ and $\gamma_4=22.98^{\circ}$. Going back to our crossed dipole interpretation of the surface admittance tensor we can imagine crossed dipoles successively rotated at each layer according to $\gamma$.  This successive rotation of each admittance surface breaks the symmetry of the transmitarray and allows for chiral behaviour as first proposed in \cite{Zhao_2012}. Again we note that we are only using the concept of surface admittances and transmission-lines without ever explicitly invoking chirality. And, as shown here, by using the procedure described above, we can design our surface to be reflectionless by properly selecting the appropriate surface admittance eigenvalues and rotation angle $\gamma$.

Simulating this unit cell in HFSS in the domain described above, we can plot the S-parameters from 9.8~GHz to 10.2~GHz as well as the fields in the unit cell.  The S-parameters are shown in Fig.~\ref{fig:SparamTwistPolRotMag} where we can see good matching in the unit cell around 10~GHz. We can also see high transmission through the cross-polarized S-parameter terms as well as the $\pi$ phase shift between the TE and TM cross polarized terms. This shows good agreement with the S-parameters in \eqref{eq:SparamPolRot}. As stated above, optimizing the bandwidth of this design is an area of future work.

Plotting the fields at 10~GHz we consider three scenarios.  When the surface is excited by a TE polarization (vertical), when the surface is excited by a TM polarization (horizontal) and when the surface is excited by a $45^{\circ}$ slanted linear polarization. This is plotted in Fig.~\ref{fig:PolRotEfieldTE}-\ref{fig:PolRotEfield45} where we can see in all three cases that the input linear polarization is rotated by $90^{\circ}$ demonstrating the ability to rotate any linear polarization.

\subsubsection{A Circular Polarizer}

\begin{figure*}[!t]
\centering
\subfloat[]
{
	\includegraphics[clip=true, trim= 0cm 4cm 0cm 3.5cm, scale=0.22]{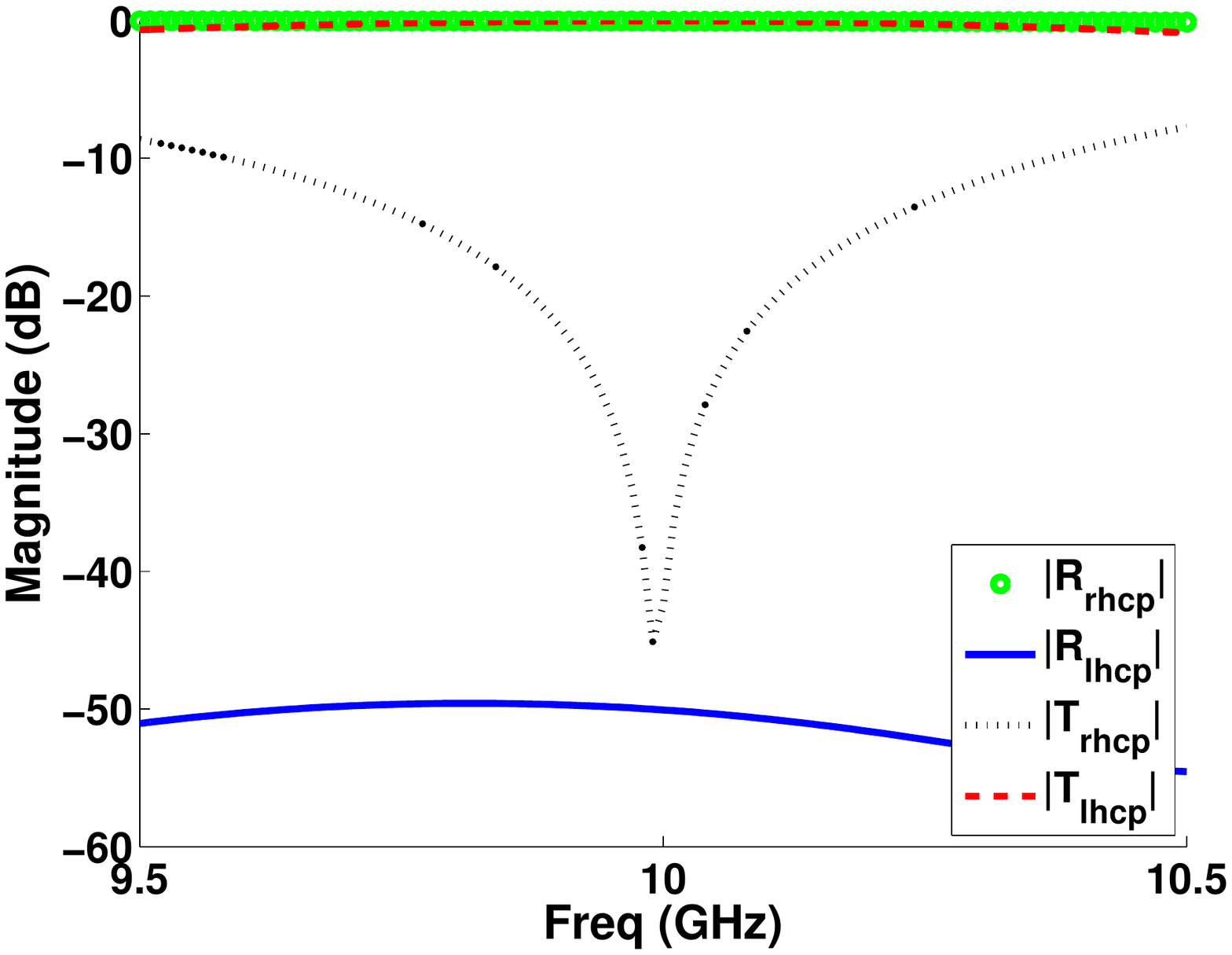}
	\label{fig:SparamCPSSMagCPBasis}
}
\subfloat[]
{
	\includegraphics[clip=true, trim=  0cm 3cm 0cm 0cm, scale=0.15]{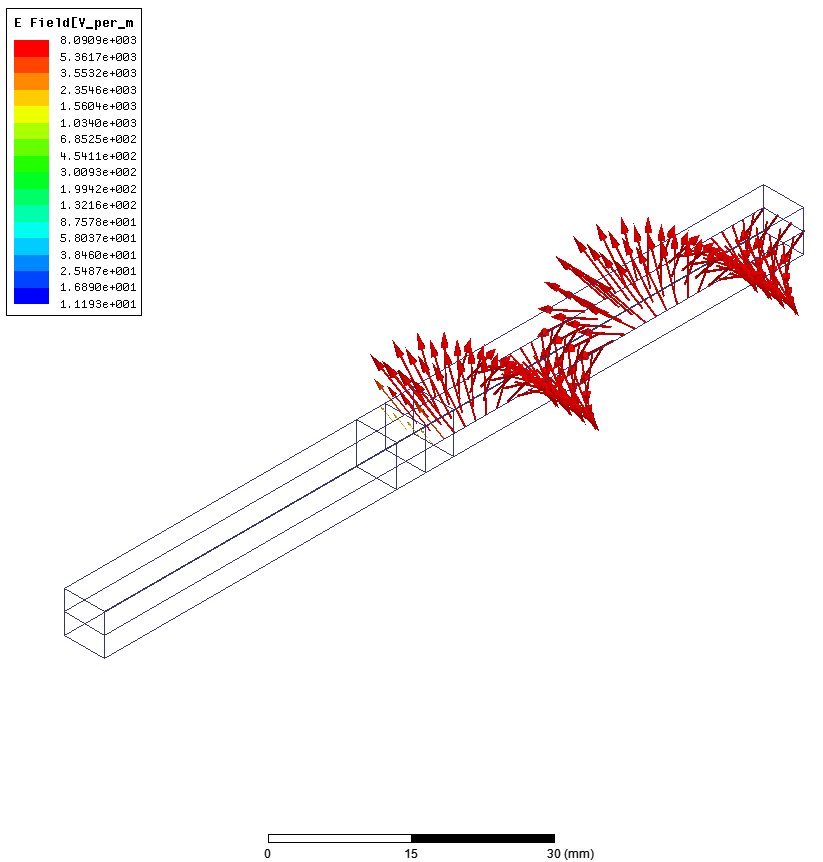}
	\label{fig:CPSSEfieldRx}
}
\subfloat[]
{
	\includegraphics[clip=true, trim=  0cm 3cm 0cm 0cm, scale=0.15]{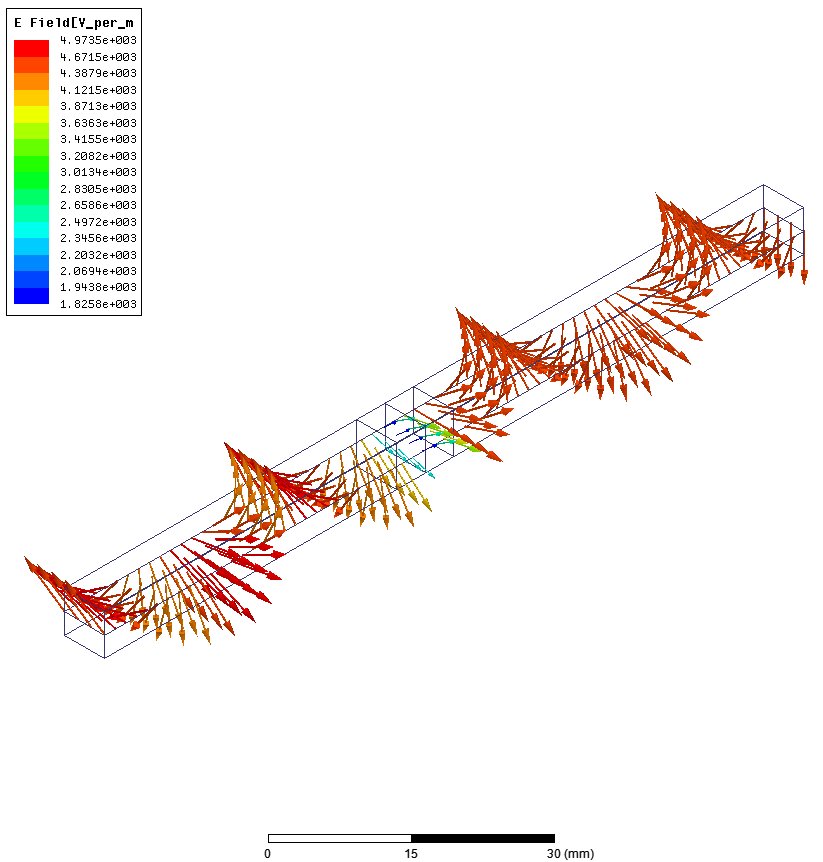}
	\label{fig:CPSSEfieldTx}
}
\caption{\protect\subref{fig:SparamCPSSMagCPBasis} The S-parameters of the circular polarizer in a circularly polarized basis. The cross coupling terms are negligible in the circular polarized basis. Note the high transmission of the LHCP wave and large reflection for the RHCP wave.  \protect\subref{fig:CPSSEfieldRx} The electric field in the unit cell when excited by a RHCP wave.  \protect\subref{fig:CPSSEfieldTx} The electric field in the unit cell when excited by a LHCP wave. }
\label{fig:HFSSCPSS}
\end{figure*}

Unlike the transmitarrays discussed so far, a circular polarizer, (also referred to a  circular polarization selective surface), is not reflectionless but is in fact designed to reflect one hand of circular polarization while passing the other \cite{Roy1996,Zhao_2012,Niemi2013}. This was implemented in \cite{Zhao_2012} using a four-layer stack of rotated nanorods with each nanorod surface acting as an admittance surface. To contrast with \cite{Zhao_2012} again, the design shown here is a design that implements the exact S-parameters of a circular polarizer for a narrow-band of frequencies while in \cite{Zhao_2012} the goal was to maximize the bandwidth by maximizing the number of layers that could be fabricated.  This is a common tradeoff in cascaded impedance structures between ripple in the transmission/reflection and bandwidth. Here, we are examining a solution at a specific frequency, trying to find the exact surface admittance of each layer to achieve such a circular polarizer.

Because our circular polarizer has reflections we do not necessarily need a four-layer design (though we could design one if desired). Instead we can get away with a three-layer design which is what we choose to do here. For a circular polarizer which rejects a right-handed circular polarization (RHCP), the S-parameters in a linearly polarized TE and TM basis are given to be \cite{Roy1996}
\begin{equation}
\label{eq:SparamCPSS}
S=\frac{1}{2} e^{-j\xi}\left[\begin{smallmatrix} -1 & 1 & j & -j  \\ 1   & -1 &  j & -j \\ j &  j  & 1 &  1 \\ -j   & -j &  1 & 1 \end{smallmatrix}\right]
\end{equation}
Thus our desired reflection coefficient matrix looking into the transmitarray is given to be
\begin{equation}
\label{eq:GammaCPSS}
\mathbf{\Gamma_{in}}=\frac{1}{2} e^{-j\xi}\left[\begin{matrix} -1 & j   \\ j   & 1 \end{matrix}\right]
\end{equation}
where the input admittance is given by inserting \eqref{eq:GammaCPSS} into \eqref{eq:YfromGamma}. Modifying the procedure given above we can simply assume a value for $\mathbf{Y_3}$ and then proceed to find $\mathbf{Y_2}$ and $\mathbf{Y_1}$ using the algebraic Riccati equation in \eqref{eq:MTLRiccati}. Here we choose to set the spacing between each admittance surface to $d=\lambda/7$.

Doing this we find an example solution where  $\mathbf{Y_1}=\left[\begin{array}{cc} -j9.63 \times 10^{-3} & -j4.14\times 10^{-3} \\ -j4.14\times 10^{-3}  & j.24\times 10^{-3} \end{array} \right]\mho$, $\mathbf{Y_2}=\left[\begin{array}{cc} -j3.27 \times 10^{-1}  & j5.69  \times 10^{-1} \\ j5.69  \times 10^{-1} & -j9.81 \times 10^{-1} \end{array} \right]\mho$ and $\mathbf{Y_3}=\left[\begin{array}{cc} -j4.78 \times 10^{-3} & -j7.03\times 10^{-3} \\ -j7.03\times 10^{-3}  & -j3.30\times 10^{-3} \end{array} \right]\mho$.  Again the value for the rotation angle of each surface is  $\gamma_1=18.66^{\circ}$, $\gamma_2=30.06^{\circ}$ and $\gamma_3=42^{\circ}$, showing the progressive twist in the rotation angle through the transmitarray.

We simulate this unit cell again in HFSS in the setup previously described and we find both the S-parameters and the fields in the structure as shown in Fig.~\ref{fig:HFSSCPSS}. The S-parameters are given in a circular polarization basis by transforming the S-parameter matrix \cite{Phillion_2011}. Here we can see in Fig.~\ref{fig:SparamCPSSMagCPBasis} a large reflection of a RHCP wave while a left-handed circularly polarized (LCHP) wave is easily transmitted by the surface.  Plotting the fields in the structure in Fig.~\ref{fig:CPSSEfieldRx} and Fig.~\ref{fig:CPSSEfieldTx} we can see that when the transmitarray is excited by a LHCP wave, it is transmitted with minimal alteration of the circular polarization state.  Correspondingly when the transmitarray is excited by a RHCP wave, the wave is reflected and very little is transmitted. 

\section{Conclusion and Future Work}
\label{sec:Conclusion}
We have demonstrated here the analysis of symmetric and asymmetric tensor impedance transmitarray made up of a stack of tensor admittance surfaces. The use of these tensor admittance sheets allows for control of both the relative amplitude and phase of the TE and TM waves that are transmitted through the array allowing for polarization control to be realized.  Using symmetric transmitarrays we have realized designs which can arbitrarily alter the polarization state for a given input polarization. This can be useful for waveplate designs as well as inhomogeneous polarization screens as shown above. For asymmetric designs we can create transmitarrays which can implement chiral behaviour such as polarization rotation and circular polarizers. We can see that this is due to the progressive twist in the rotation angles of each surface admittance tensor in the transmitarray. 

Both these symmetric and asymmetric designs were analyzed using a MTL model which treats each surface admittance as a shunt load on a $2+1$ wire MTL line. Doing this allows us to use concepts such as the input admittance and reflection coefficient to design the transmitarrays  to be reflectionless.

The design techniques laid out in this paper can be used to further explore possible solutions for controlling the polarization and wavefronts of an incident wave. This can include varying the dielectric spacers between the surface admittance, introducing loss (judiciously) and looking a surfaces with small reflections to create novel designs. 

With regards to their implementation, various designs of tensor admittances surfaces exist at both microwave and optical frequency beyond the crossed dipole. At microwaves, these designs include diagonal slots embedded in patches \cite{Fong_etal_2010}, diagonal metallic patterns \cite{Patel_2013} and rotated or skewed dipoles \cite{Zhao_etal_2011,Selvanayagam2014}. At optical frequencies, tensor impedance surfaces have been fabricated out of gold nanorods \cite{Zhao_2012,Grady2013} and `V' antennas \cite{Yu_2012}. The analytical techniques presented here can help to further optimize the single layer and stacked structures fabricated in \cite{Yu_2012,Zhao_2012,Grady2013}   


Thus, the analysis of a tensor impedance transmitarray presented here offers a way at both microwave and optical frequencies to implement planar reflectionless designs capable of polarization control.

\section*{Appendix A - S-parameters Of A Tensor Impedance Transmitarray}
For a symmetric or asymmetric tensor impedance transmitarray we use the techniques given above to find a solution for the tensor surface admittances that yields a reflectionless design. With this solution we would like to evaluate the transmission through the layered admittance also.  To do this we use $4\times4$ matrices to define the relationship between the network ports, with 2 ports for the input and 2 ports for the output (one for each polarization respectively). Each component of the transmitarray can be represented by a $4\times4$  network matrix. An example of a four-port network for a shunt surface admittance was shown in Fig.~\ref{fig:ShuntYMTL} with the corresponding port numberings.  For a shunt admittance sheet on a transmission-line a $4\times4$ impedance matrix representation was defined in \cite{Selvanayagam2014} and is given by,
\begin{align}
\mathbf{Z_{sh}}=\left[\begin{array}{cccc} Z_{yy}  &  Z_{yy} & Z_{yz} & Z_{yz} \\
 Z_{yy}  &  Z_{yy} & Z_{yz} & Z_{yz} \\
Z_{zy}  &  Z_{zy} & Z_{zz} & Z_{zz} \\
 Z_{zy}  &  Z_{zy} & Z_{zz} & Z_{zz} 
\end{array}\right] 
\end{align}
where the elements of $\mathbf{Z_{sh}}$ are found from $\mathbf{Z_{\{1-4\}}}=\mathbf{Y^{-1}_{\{1-4\}}}$. This impedance matrix can be converted to an S-parameter matrix using $\mathbf{S}=(\mathbf{Z_{sh}}-\mathbf{Z_o})(\mathbf{Z_{sh}}+\mathbf{Z_o})^{-1}$, where $\mathbf{Z_o}$ is a diagonal matrix of the port impedance of each port \cite{Pozar}.   In our case the port impedance at each port is simply $\eta$, the free space wave impedance. 

For the interconnecting transmission-lines of length $d$ and characteristic impedance $\eta_1$ and propagation constant $k_o$ which model free space, the TE and TM modes are decoupled as stated earlier.  Thus the S-parameter matrix is simply given below by \eqref{eq:STL}.
\begin{figure*}[!t]
\begin{align}
\label{eq:STL}
\mathbf{S_{TL}}=\left[\begin{smallmatrix}
\frac{j\sin k_o d (\eta_1-\eta^2/\eta_1)}{2\eta\cos k_o d+j(\eta_1+\eta^2/\eta_1)\sin k_od}  &  \frac{2\eta}{2\eta\cos k_o d+j(\eta_1+\eta^2/\eta_1)\sin k_od} & 0 & 0 \\
 \frac{2\eta}{2\eta\cos k_o d+j(\eta_1+\eta^2/\eta_1)\sin k_od}   & \frac{j\sin k_o d (\eta_1-\eta^2/\eta_1)}{2\eta\cos k_o d+j(\eta_1+\eta^2/\eta_1)\sin k_od}  & 0 & 0 \\
0  &  0 & \frac{j\sin k_o d (\eta_1-\eta^2/\eta_1)}{2\eta\cos k_o d+j(\eta_1+\eta^2/\eta_1)\sin k_od}& \frac{2\eta}{2\eta\cos k_o d+j(\eta_1+\eta^2/\eta_1)\sin k_od}  \\
0  &  0 &\frac{2\eta}{2\eta\cos k_o d+j(\eta_1+\eta^2/\eta_1)\sin k_od} &\frac{j\sin k_o d (\eta_1-\eta^2/\eta_1)}{2\eta\cos k_o d+j(\eta_1+\eta^2/\eta_1)\sin k_od}
\end{smallmatrix}\right] 
\end{align}
\hrulefill
\hrulefill
\end{figure*}

To determine the S-matrix of the overall transmitarray we can convert all the S-matrices into $4\times4$ transfer matrices which can be multiplied together to model the stack of admittance sheets and dielectric layers as given by,
\begin{align}
\label{eq:Tarray}
\mathbf{T_{array}}=\mathbf{T_{Y,1}}\mathbf{T_{TL}}\mathbf{T_{Y,2}} ...\mathbf{T_{TL}}\mathbf{T_{Y,N-1}}\mathbf{T_{TL}}\mathbf{T_{Y,N}}.
\end{align}
 This overall transfer matrix can be converted back to an S-parameter matrix to find the transmission through the transmitarray.   The procedure for converting between S-parameters and transfer matrices is given in \cite[Appendix C]{Lau}.

Note that it is technically possible to solve for all the required admittances of the transmitarray by using the total transfer matrix in \eqref{eq:Tarray}. However this would require rearranging the set of equations into a more useful form which would be very difficult due to the number of matrix inversions required to convert S and Z parameters to transfer matricies. Hence, our choice of using a semi-analytical method to find the surface admittances.
 

\end{document}